\newif\ifsingle
\singletrue
\ifsingle
\documentclass[12pt, journal, onecolumn, draftclsnofoot]{IEEEtran}
\else
\documentclass[10pt, johttps://www.overleaf.com/project/5d8b8abcbc45c7000194efbeurnal,letterpaper]{IEEEtran}
\fi

\IEEEoverridecommandlockouts

\usepackage{algorithm,amsmath,amssymb,amsthm,array,bbm,bibentry,cite,color,bbm}
\usepackage{enumerate,enumitem,eurosym,float,graphicx,mathrsfs,mathtools,multirow,nomencl}
\usepackage{pict2e,psfrag,ragged2e,setspace,stfloats,supertabular,tabularx,url,algorithmicx}
\usepackage{algpseudocode}
\usepackage[USenglish]{babel}
\usepackage{array}
\newcolumntype{C}[1]{>{\centering\arraybackslash}m{#1}}
\usepackage{adjustbox}
\usepackage{comment}
\usepackage[utf8]{inputenc}
\usepackage[T1]{fontenc}
\usepackage[table]{xcolor}
\usepackage{xcolor,etoolbox}
\DeclarePairedDelimiter\abs{\lvert}{\rvert}%
\DeclarePairedDelimiter\norm{\lVert}{\rVert}%

{}
{}
{}
{}
{}
{}
{}
\newtheorem{problem}{Problem}

\newcommand{\rmL}{\mathrm{d}}
\newcommand{\Lc}{\mathcal{L}}
\newcommand{\Los}{\mathrm{LoS}}
\newcommand{\Nlos}{\mathrm{NLoS}}
\newcommand{\rmR}{\mathrm{R}}
\newcommand{\mrt}{\textnormal{\tiny{MRT}}}
\newcommand{\rmUE}{\textnormal{\tiny{UE}}}
\newcommand{\minimize}{\mathrm{minimize}}

\newcommand{\name}{RISMA}
\makeatletter 
\pretocmd\@bibitem{\color{black}\csname keycolor#1\endcsname}{}{\fail}
\makeatother

\newcommand{\st}{\mathrm{subject~to}}

\makeatletter
\let\oldabs\abs
\def\abs{\@ifstar{\oldabs}{\oldabs*}}

\let\oldnorm\norm
\def\norm{\@ifstar{\oldnorm}{\oldnorm*}}
\makeatother
\renewcommand{\a}{\mathbf{a}}
\renewcommand{\b}{\mathbf{b}}

\newcommand{\e}{\mathbf{e}}

\newcommand{\g}{\mathbf{g}}
\newcommand{\h}{\mathbf{h}}

\newcommand{\s}{\mathbf{s}}

\renewcommand{\v}{\mathbf{v}}
\newcommand{\w}{\mathbf{w}}
\newcommand{\x}{\mathbf{x}}

\newcommand{\z}{\mathbf{z}}

\newcommand{\0}{\mathbf{0}}
\newcommand{\1}{\mathbf{1}}

\newcommand{\B}{\mathbf{B}}

\newcommand{\G}{\mathbf{G}}
\renewcommand{\H}{\mathbf{H}}
\newcommand{\I}{\mathbf{I}}

\newcommand{\Q}{\mathbf{Q}}

\newcommand{\V}{\mathbf{V}}
\newcommand{\W}{\mathbf{W}}
\newcommand{\X}{\mathbf{X}}

\newcommand{\mub}{\boldsymbol{\mu}}

\newcommand{\Phib}{\mathbf{\Phi}}

\newcommand{\Real}{\mbox{$\mathbb{R}$}}
\newcommand{\Compl}{\mbox{$\mathbb{C}$}}

\newcommand{\argmin}{\operatornamewithlimits{argmin}}

\newcommand{\diag}{\mathrm{diag}}

\newcommand{\Exp}{\mathbb{E}}

\newcommand{\herm}{\mathrm{H}}
\renewcommand{\Im}{\mathrm{Im}}

\renewcommand{\Re}{\mathrm{Re}}
\newcommand{\tr}{\mathrm{tr}}
\newcommand{\tran}{\mathrm{T}}

\title{\name: Reconfigurable Intelligent Surfaces Enabling Beamforming for IoT Massive Access}
\ifsingle
\author{Placido Mursia,~\IEEEmembership{Student Member,~IEEE,}
	    Vincenzo~Sciancalepore,~\IEEEmembership{Senior Member,~IEEE,}
	    Andres Garcia-Saavedra,
        Laura Cottatellucci,~\IEEEmembership{Member,~IEEE},
        Xavier~Costa-P\'erez,~\IEEEmembership{Senior Member,~IEEE}
        and~David Gesbert,~\IEEEmembership{Fellow,~IEEE}
\thanks{\textit{P. Mursia and D. Gesbert are with Communication Systems Department, EURECOM, Sophia-Antipolis, France. Emails: \{placido.mursia, david.gesbert\}@eurecom.fr}}%
	\thanks{\textit{V. Sciancalepore, A. Garcia-Saavedra and X. Costa-P\'erez are with NEC Laboratories Europe, 69115 Heidelberg, Germany. Emails: \{vincenzo.sciancalepore, andres.garcia.saavedra, xavier.costa\}@neclab.eu}}%
	\thanks{\textit{L. Cottatellucci is with Institute for Digital Communications, University of Erlangen-Nürnberg , Erlangen, Germany. Email: laura.cottatellucci@fau.de}}
}
\else
\author{Placido Mursia,~\IEEEmembership{Student Member,~IEEE,}
	    Vincenzo~Sciancalepore,~\IEEEmembership{Senior Member,~IEEE,}
	    Andres Garcia-Saavedra,\\
        Laura Cottatellucci,~\IEEEmembership{Member,~IEEE},
        Xavier~Costa-P\'erez,~\IEEEmembership{Senior Member,~IEEE}\\
        and~David Gesbert,~\IEEEmembership{Fellow,~IEEE}
\thanks{\textit{P. Mursia and D. Gesbert are with Communication Systems Department, EURECOM, Sophia-Antipolis, France. Emails: \{placido.mursia, david.gesbert\}@eurecom.fr}}%
	\thanks{\textit{V. Sciancalepore, A. Garcia-Saavedra and X. Costa-P\'erez are with NEC Laboratories Europe, 69115 Heidelberg, Germany. Emails: \{vincenzo.sciancalepore, andres.garcia.saavedra, xavier.costa\}@neclab.eu}}%
	\thanks{\textit{L. Cottatellucci is with Institute for Digital Communications, University of Erlangen-Nürnberg , Erlangen, Germany. Email: laura.cottatellucci@fau.de}}
}
\fi

\begin{document}
\bstctlcite{IEEEexample:BSTcontrol}
\maketitle

\setlength{\textfloatsep}{5pt}

\thispagestyle{empty}	

\begin{abstract}
Massive access for Internet-of-Things (IoT) in beyond 5G networks represents a daunting challenge for conventional bandwidth-limited technologies. Millimeter-wave technologies (mmWave)---which provide large chunks of bandwidth at the cost of more complex wireless processors in harsher radio environments---is a promising alternative 
to accommodate massive IoT but its cost and power requirements are an obstacle for wide adoption in practice. 
In this context,  meta-materials arise as a key innovation enabler to address this challenge by Re-configurable Intelligent Surfaces (RISs).


In this paper we take on the challenge and study a beyond 5G scenario consisting of a multi-antenna base station (BS) serving a large set of single-antenna user equipments (UEs) with the aid of RISs to cope with non-line-of-sight paths. Specifically, we build a  mathematical framework to jointly optimize the precoding strategy of the BS and the RIS parameters in order to minimize the system sum mean squared error (SMSE). This novel approach reveals convenient properties used to design two algorithms, \name{} and Lo-\name{}, which are able to either find simple and efficient solutions to our problem (the former) or accommodate practical constraints with low-resolution RISs (the latter). Numerical results show that our algorithms outperform conventional benchmarks that do not employ RIS (even with low-resolution meta-surfaces) with gains that span from $20\%$ to $120\%$ in sum rate performance. 
\end{abstract}
\begin{IEEEkeywords}
 Massive access, mmWave, Re-configurable intelligent surfaces, Beamforming, IoT, Beyond 5G.
\end{IEEEkeywords}
 
\section{Introduction}\label{sec:introduction}

Spurred by economic and environmental concerns, the design of energy-efficient high-bandwidth wireless technologies is becoming paramount---even small improvements matter at the scale of next-generation Internet-of-Things (IoT) systems~\cite{buzzi2016survey}. 
We argue in this paper that a joint exploitation of  \emph{millimeter-wave}  spectrum (mmWave), which can provide multi-GHz bandwidth, \emph{and} \emph{Re-configurable Intelligent Surfaces} (RISs),\footnote{Note that the term \emph{Intelligent} \emph{Reflecting} \emph{Surface} (IRS) is alternatively used in other related work. The acronyms IRS and RIS can be used interchangeably as they refer to the same physical device, i.e., a reflecting surface that can be controlled by some network entity. To avoid ambiguity we have chosen to use the term RIS.} which can alleviate the energy toll attained to the former, has the potential to achieve this goal. 

{\bf RISs: aiding and abetting Massive IoT access based on mmWave technology.} 
The hunt for wider radio bands has led network practitioners to study, with success, the use of mmWave as a means to accommodate broadband connectivity. In fact, mmWave is doubtlessly one of the key building stones of 5G and will continue to be so in future-generation systems. 
However, the low-power low-throughput nature of conventionally-deployed IoT devices have caused such high-frequency bands, with considerably harsher propagation properties, to be largely ignored when building IoT environments. Nevertheless, the advent of massive IoT applications spawning a huge volume of devices puts a strain on low-bandwidth sub-6GHz technologies and poses mmWave as a candidate solution for quasi-nomadic scenarios such as smart grids, smart cities and smart industries~\cite{8570917}. The main challenge in this case is that mmWave transceivers usually employ digital or hybrid beamforming, with multiple RF chains and a large number of antenna arrays that allow focusing electromagnetic energy into certain angles (i.e., irradiate beams), in order to combat mmWave's \emph{aquaphobia} and high attenuation. This strategy is however doomed for energy-constrained IoT devices, as integrating multiple active components draining energy becomes infeasible~\cite{8850102}. Faced with such a challenge, \emph{RISs may hold the key to properly exploiting the use of mmWave with its vast bandwidth resources while enabling advanced massive IoT scenarios with a significantly-low service disruption probability \cite{Ren20}.}

Indeed, RISs, which apply controllable transformations into impinging radio waves {without leveraging on power amplifiers}, create a host of opportunities for the optimization of wireless systems at a low cost and with a low energy footprint~\cite{Ren19}. They are in fact gaining a lot of momentum~\cite{Sub12, Tan16, Zha19, Bas19, Bas19_2, Ozd19, Tan19, Hu18, Zha19_2, Cui19, Yu19_2, Fu19, Jia19, Hua19, Hua19_2, Hua19_3} because of their ability to turn the stochastic nature of the wireless environment---fundamentally passive---into a programmable channel that plays an active role on the way in which signals propagate. RISs have been recently proposed for a variety of applications, ranging from secure communications~\cite{Cui19, Yu19_2}, non-orthogonal multiple access~\cite{Fu19}, over-the-air-computation~\cite{Jia19} or energy-efficient cellular networks~\cite{Hua19, Hua19_2}.
A RIS is essentially a continuous meta-surface that can be modeled as a grid of discrete unit cells spaced at sub-wavelength distance that can alter their electromagnetic response, such as phase, amplitude, polarization and frequency in a programmable manner. For instance, they can be tuned such that signals bouncing off a RIS  are combined constructively to increase signal quality at the intended receiver or destructively to avoid leaking signals to undesired receivers.

{\bf RISs \emph{vs.} relaying and MIMO.}
Conceptually, a RIS may remind some of the challenges behind conventional \emph{Amplify-and-Forward} (AF) relaying methods~\cite{Nto19} and the beamforming methods used in (massive) MIMO~\cite{bjrnson2019demystifying}. 
There exists a marked difference between conventional AF relays and RISs \cite{Nto19_2, Bjo19}. Indeed, the former rely on active (energy-consuming) low-noise power amplifiers and other active electronic components, such as digital-to-analog (DAC) or analog-to-digital (ADC) converters, mixers and filters. In contrast, RISs have very low hardware footprint, consisting on a single or just a few layers of planar structures that can be built using lithography or nano-printing methods. Consequently, RISs result to be particularly attractive for seamless integration into walls, ceilings, object cases, building glasses or even clothing~\cite{Falade_2018}. 

On the other hand, (massive) MIMO employs a large number of antennas to attain large beamforming gains. In fact, upon similar conditions, both massive MIMO and RIS technology can produce similar signal-to-noise-ratio (SNR) gains.\footnote{Although it has been shown that the SNR scales linearly with the number of antennas $M$ when using massive MIMO and proportional to the square of the number of equivalent antenna elements $N^2$ with RIS technology, the lack of power amplification in the latter determines a performance loss such that, overall, both technologies produce very similar SNR gains given the same conditions~\cite{bjrnson2019demystifying}.} However, a RIS achieves such beamforming gains passively---with a negligible power supply---exhibiting high energy efficiency. We claim in this paper that active beamforming via an antenna array at the transmitter side and passive beamforming in the channel via a RIS \emph{can complement each other and provide even larger gains  when they both are jointly optimized, which is precisely the goal of this paper.} To this aim---and in marked contrast to earlier works~\cite{Yu19, Nad19, Wu18, Mis19, Wu19, Kar19, Hua20}---we use the received sum mean squared error (SMSE) as optimization objective, which let us \emph{find simple and efficient solutions for the problem at hand}.

While the theoretical modelling of RIS-aided wireless networks is well studied, many challenges are still open to be tackled such as building testbeds for experimental validation \cite{Tan19_2, Aru19, Liu19, Dai19}, the task of estimating the combined channel from the BS to the RIS and on to the UE \cite{He19, Mis19} and the joint optimization of the multi-antenna BS and the RIS parameters \cite{Yu19, Nad19, Wu18, Mis19, Wu19, Kar19, Hua20}. Particularly relevant for this paper is the latter category, concerning the joint optimization of active beamforming at the BS and passive beamforming at the RIS. In \cite{Yu19} the authors analyze a single-UE case and propose to maximize the rate. The resulting non-convex optimization problem is solved via both fixed point iteration and manifold optimization. A similar setting is analyzed in \cite{Mis19}, where the authors propose a heuristic solution to the non-convex maximization of the received signal power with similar performance to conventional semidefinite relaxation (SDR). The single-UE setting is also studied in \cite{Kar19}, where the authors propose to encode information both in the transmitted signal and in the RIS configuration. A multiuser setting is analyzed in \cite{Nad19} where the authors propose to maximize the minimum receive SNR among all UEs in the large system regime. While this approach guarantees fairness among UEs, it might not maximize the system sum rate. In \cite{Wu18}, the authors design jointly the beamforming at the BS side and the RIS parameters by minimizing the total transmit power at the BS, given a minimum receive signal-to-interference-plus-noise ratio (SINR) requirement. This framework was later extended to consider low resolution RISs in a single UE setting \cite{Wu19}.

\subsection{Novelty and contributions}

The main novelty of this paper stems from exploiting the SMSE as an optimization objective. The choice of an objective function is of paramount importance, especially for massive access scenarios. Our objective function is purposely chosen such that we can derive a mechanism that provides high-performing solutions while guaranteeing efficiency and scalability. Interestingly, such metric---which has not been studied so far in the context of RIS-aided networks---reveals a convex structure in the two optimization variables separately, namely the precoding strategy at the transmitter and the RIS parameters. This gives us an edge over prior work because it allows to design very efficient iterative algorithms for RIS control. Specifically, we present \name{}, a RIS-aided Multiuser Alternating optimization algorithm that jointly optimizes the beamforming strategy at the transmitter (a BS) and the RIS parameters to provide high-bandwidth low-cost connectivity in massive IoT scenarios. In marked contrast with prior work, \name{} exploits the convex nature of the problem at hand in the two optimization variables separately to ensure scalability, efficiency and provable convergence in the design without the need of setting any system parameter.

Moreover, we adapt \name{}, which provides a solution from a theoretical perspective, to accommodate practical constraints when using low-resolution RISs that are comprised of antenna elements that can be activated in a binary fashion. In this way, these are meta-surfaces that only support phase shift values from a discrete set, rather than any real value from a range, and further compound our problem \cite{Hum14,Men16}. To address this scenario, we propose Lo-\name{}, which decouples the optimization of the binary activation coefficients and the quantized phase shits. The former are optimized via SDR while the latter are projected onto the quantized space. {Differently than other prior work considering low-resolution RISs \cite{Wu19, Di20}, Lo-\name{} benefits from the key properties of the chosen SMSE metric. Specifically,  for  each  iteration  of  the  proposed algorithm for a fixed RIS configuration the precoding strategy is found via a simple closed form solution. Whereas once  the  precoding  strategy  is  fixed  the  problem  of  finding  the  RIS  parameters can be efficiently solved via SDR.}

Our numerical results show that a joint optimization of both the precoder of the transmitter and the RIS parameters in terms of induced phase shifts and amplitude attenuation produce substantial gains in sum rate performance. Specifically, our joint optimization approach leads to $\sim$40\% gain compared to using only a minimum mean squared error (MMSE) precoder over a broad range of network area radii, and gains that scale linearly with the network-area's radius compared to a zero-forcing (ZF) precoder, e.g., $\sim$20\% and $\sim$120\% improvement for radii equal to $100$ and $150$ meters, respectively.

To summarize, the contributions of this paper are:
\begin{itemize}
    \item We introduce a novel mathematical framework to minimize the SMSE of RIS-aided beamforming communication systems that make them suitable for massive IoT wireless access. This approach, which to the best of our knowledge has not been explored before, allows us to build efficient algorithms that maximize sum rate performance.
    \item We design \name{}, a low-complexity scheme with provable convergence that finds a simple and efficient solution to the aforementioned problem.
    \item We introduce Lo-\name{}, an efficient algorithm for realistic scenarios with low-resolution meta-surfaces.
    \item We present a thorough numerical evaluation that shows substantial gains in terms of sum rate performance. Specifically, we present scenarios where our approach achieves around $40\%$ gain over an MMSE precoder, and gains that span between $20\%$ and over $120\%$ with respect to a ZF precoder, depending on the network radius.
\end{itemize}

The remaining of this paper is structured as follows. Section~\ref{sec:sys_model} introduces the system model and the problem formulation to optimize the considered metric. In Section~\ref{sec:singleUE} we tackle the solution of the considered problem in the simple case of a single UE. In Section~\ref{sec:multiUE} the aforementioned problem is solved through the proposed \name{} algorithm in a general multiuser setting. Moreover, we propose Lo-\name{} which provides a practical implementation of \name{} in the case of low-resolution RIS. Section~\ref{sec:Results} presents numerical results to evaluate the performance of the proposed algorithms. Lastly, Section~\ref{sec:concl} concludes the paper.

\subsection{Notation} 
Throughout the paper, we use italic letters to denote scalars, whereas vectors and matrices are denoted by bold-face lower-case and upper-case letters, respectively. We let $\Compl$, $\Real$ and $\mathbb{Z}$ denote the set of complex, real and integer numbers, respectively. We use $\Compl^n$ and $\Compl^{n\times m}$ to represent the sets of $n$-dimensional complex vectors and  $m\times n$ complex matrices, respectively. Vectors are denoted by default as column vectors. Subscripts represent an element in a vector and superscripts elements in a sequence. For instance, $\X^{(t)} =   [x^{(t)}_1, \dots,  x^{(t)}_n]^{\tran}$ is a vector from $\Compl^n$ and  $x^{(t)}_i$ is its $i$th component. Operation $(\cdot)^{\tran}$ represents the transpose operator, $\otimes$ stands for Kronecker product while $(\cdot)^{\herm}$ denotes the Hermitian transpose operation. Finally, $\|\cdot
\|$ and $\|\cdot\|_{\mathrm{F}}$ denote the L2-norm of a vector and Frobenius norm of a matrix, respectively.


\section{Model design}\label{sec:sys_model}

\subsection{System model}\label{subsec:sys_model}
\ifsingle
\begin{figure}[t!]
    \centering
    \includegraphics[width=0.6\columnwidth, clip]{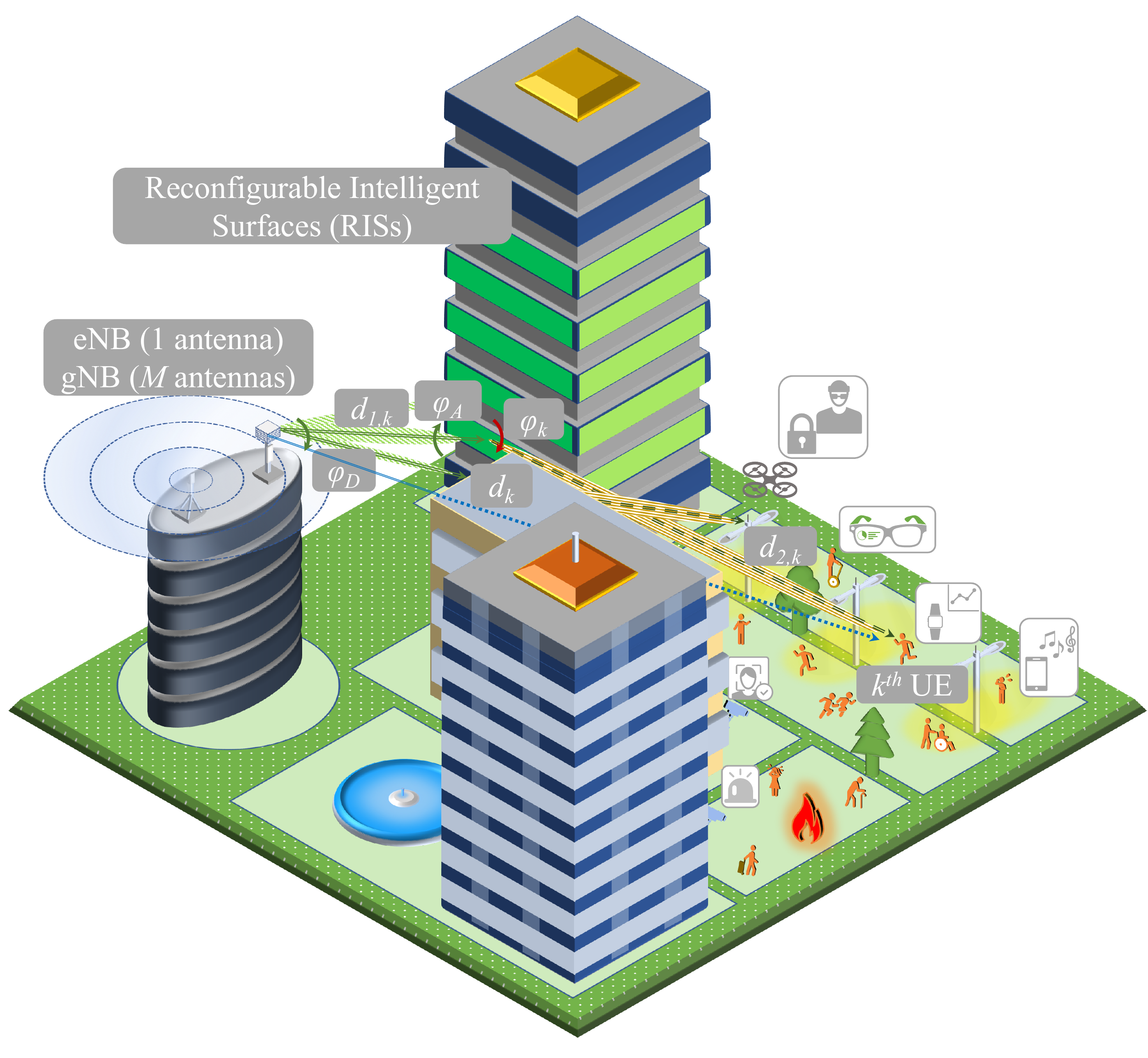}\vspace{-4mm}
    \caption{Radio massive access scenario overcoming NLOS issues by means of RISs installed on the building glasses. It might support different use cases, such as AR-glasses, e-health, video-survelliance, Industrial-IoT.}
    \label{fig:scenario}
\end{figure}
\else
\begin{figure}[t!]
    \centering
    \includegraphics[width=\columnwidth, clip]{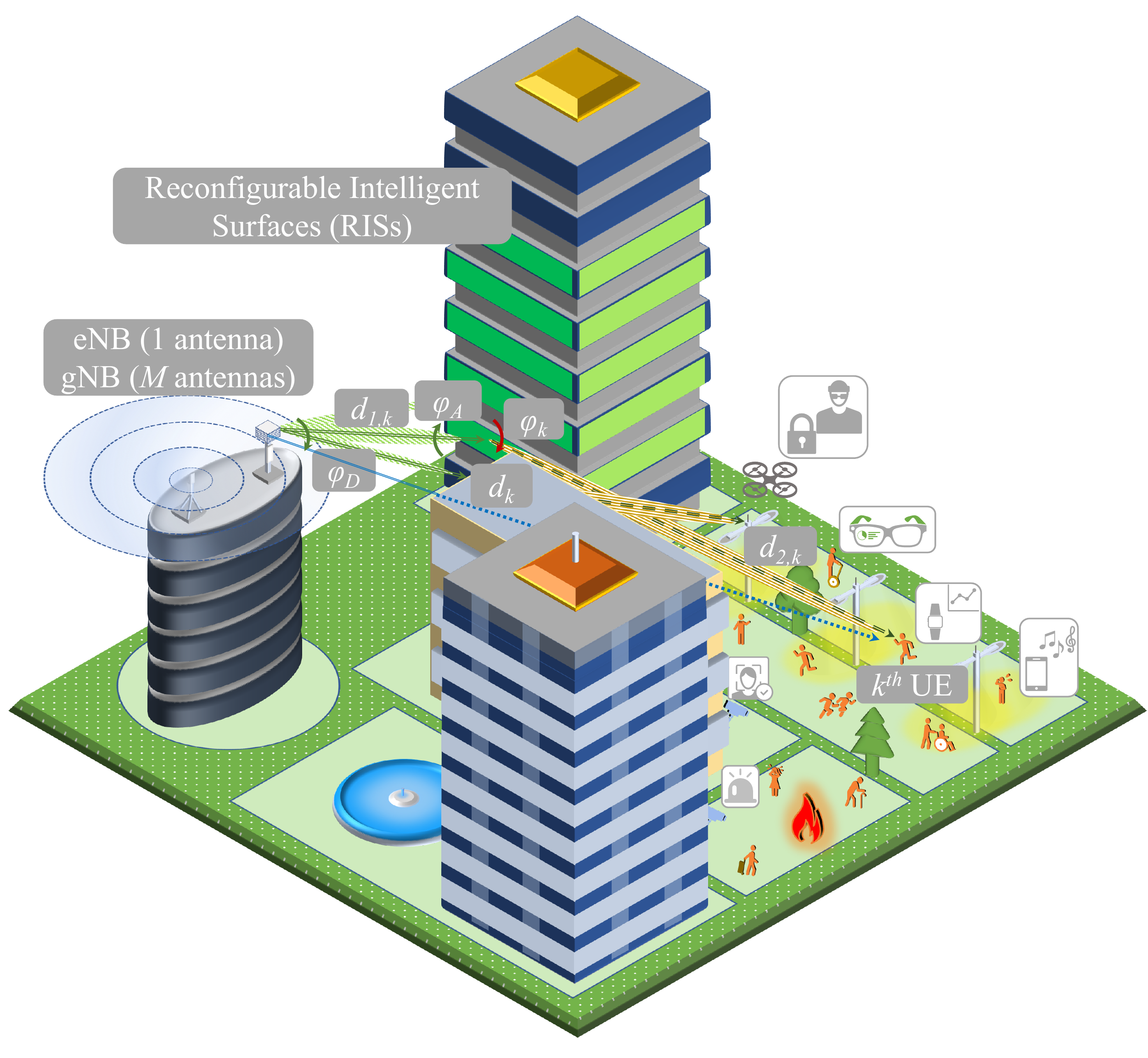}\vspace{-4mm}
    \caption{Radio massive access scenario overcoming NLOS issues by means of RISs installed on the building glasses. It might support different use cases, such as AR-glasses, e-health, video-survelliance, Industrial-IoT.}
    \label{fig:scenario}
\end{figure}
\fi

Let us consider the scenario described in Fig.~\ref{fig:scenario} in which a BS equipped with $M$ antennas serves a set of $K$ single-antenna user equipment nodes (UEs).
 However, note that the proposed method is not limited to such a case. When considering multiple-antenna UEs, our model can be readily applied by letting each UE activate the antenna with the highest average channel power gain. The connection is established with the aid of a set of RISs installed on the building glasses each of which consists of $N$ equivalent antenna elements. Focusing on the downlink data transmission, the BS communicates to each UE~$k$ via a direct link denoted by $\h_{\rmL,k} \in \Compl^{M\times 1}$ which comprises of a line-of-sight (LoS) path of length $d_k$ and angle of departure (AoD) $\theta_k$ when the latter exists, in addition to a multipath non-line-of-sight (NloS) link. Additionally, the BS can exploit a combined link from the BS to the RIS denoted by $\G \in \Compl^{N\times M}$, which in turns reflects the incoming signal towards the UE through the channel $\h_k \in \Compl^{N\times 1}$. The latter is decomposed into the LoS BS-RIS path of length $d_{1,k}$ and with AoD from the BS and angle of arrival (AoA) at the RIS denoted by $\psi_{D}$ and $\psi_{A}$, respectively plus a set of scarreted NLos paths and the RIS-UE~$k$ link which comprises of a LoS path of length $d_{2,k}$ and AoD $\psi_k$ when available, plus a multipath NLoS link. Lastly, due to high path loss we neglect all signals reflected two times or more by the RIS as in \cite{Zha19_2, Wu18, Mis19}.
 
All channels follow a quasi-static flat-fading model and thus remain constant over the transmission time of a codeword. We further assume that perfect channel state information (CSI) is available at the BS, i.e., the latter knows $\{\h_{\rmL,k}\}_{k=1}^K$, $\G$ and $\{\h_{k}\}_{k=1}^K$. The BS operates in time-division duplexing mode, such that the uplink and downlink channels are reciprocal. The downlink physical channel can thus be estimated through the uplink training from the UEs via a separate control channel\,\footnote{When dealing with biased channel information, a channel estimation process is required. However, such a challenge in RIS-aided networks has been already explored in~\cite{He19, Mis19} and thus is out of the scope of this work.}.

While we focus on the downlink data transmission, our proposed framework might be straightforwardly extended to the uplink direction considering multiple UEs and one single BS. Each UE~$k$ receives the sum of two contributions, namely a direct path from the BS and a suitably reflected path upon the RIS. Hence, the receive signal at UE~$k$ is given by
\ifsingle
\begin{align}
y_k & = \left(\h_k^{\herm}\Phib\G + \h_{\rmL,k}^{\herm}\right)\W\s + n_k \in \Compl \\
& = \left(\h_k^{\herm}\Phib\G + \h_{\rmL,k}^{\herm}\right)\w_k s_k + \sum_{j\neq k} \left(\h_k^{\herm}\Phib\G + \h_{\rmL,k}^{\herm}\right)\w_j s_j + n_k,
\end{align}
\else
\begin{align}
y_k & = \left(\h_k^{\herm}\Phib\G + \h_{\rmL,k}^{\herm}\right)\W\s + n_k \in \Compl \\
& = \left(\h_k^{\herm}\Phib\G + \h_{\rmL,k}^{\herm}\right)\w_k s_k \nonumber \\
& \phantom{= } + \sum_{j\neq k} \left(\h_k^{\herm}\Phib\G + \h_{\rmL,k}^{\herm}\right)\w_j s_j + n_k,
\end{align}
\fi
where  $\Phib = \mathrm{diag}[\alpha_1e^{j\phi_1},\ldots,\alpha_Ne^{j\phi_N}]$ with $\phi_i \in [0,2\pi)$ and $|\alpha_i|^2\leq 1, \,\, \forall i$ represents the phase shifts and amplitude attenuation introduced by the RIS (\cite{Bas19,Hua18,Wu18,Ozd19}), $\W = [\w_1,\ldots,\w_K] \in \Compl^{M\times K}$ is the transmit precoder, $\s = [s_1,\ldots,s_K]^{\tran}\in \Compl^{K\times 1}$ is the transmit symbol vector with $\Exp[|s_k|^2] = 1, \, \forall k$, and $n_k$ is the noise term distributed as $\mathcal{CN}(0,\sigma_n^2)$. 

Hence, assuming single-user decoding at the receiver side the system sum rate can be defined as follows
\begin{align}\label{eq:srate}
    R\! \triangleq \! \sum_k \log_2\bigg(\! 1\!+\! \frac{|(\h_k^{\herm}\Phib\G + \h_{\rmL,k}^{\herm})\w_k|^2}{\sum_{j\neq k} |(\h_k^{\herm}\Phib\G + \h_{\rmL,k}^{\herm})\w_j|^2 + \sigma_n^2}\bigg).
\end{align}

\subsection{Problem Formulation}\label{subsec:problem}

Our objective is to optimize the overall system performance of the considered RIS-aided network in terms of the system sum rate, as defined in Eq.~\eqref{eq:srate}. In particular, given the complexity of treating such an expression, we propose to jointly optimize the precoding strategy at the BS and the reflections (as a tunable parameter) introduced by the RIS by minimizing the SMSE over all connected UEs, which is known to relate to the sum rate~\cite{Chr08}. In particular, for a given configuration of the RIS the considered system in the downlink is a broadcast channel and duality between broadcast and uplink multiple access channel holds. In the dual multiple access channel the classical relation between minimum mean squared error (MSE) of UE $k$ and maximum SINR of UE $k$ holds for linear filters \cite{Hun09}. Hence, this motivates us to study the SMSE as a means to optimize the system sum rate in the downlink.

The receive MSE of UE~$k$ is given by
\ifsingle
\begin{align}
\mathrm{MSE}_k & = \Exp[|y_k - s_k|^2]\\
& = |(\h_k^{\herm}\Phib\G + \h_{\rmL,k}^{\herm})\w_k - 1|^2 + \sum_{j\neq k} |(\h_k^{\herm}\Phib\G + \h_{\rmL,k}^{\herm})\w_j|^2 + \sigma_n^2 \\
& = \sum_j |(\h_k^{\herm}\Phib\G + \h_{\rmL,k}^{\herm})\w_j|^2 - 2 \, \Re\{(\h_k^{\herm}\Phib\G + \h_{\rmL,k}^{\herm})\w_k\} + 1 + \sigma_n^2.
\end{align}
\else
\begin{align}
\mathrm{MSE}_k & = \Exp[|y_k - s_k|^2]\\
& = |(\h_k^{\herm}\Phib\G + \h_{\rmL,k}^{\herm})\w_k - 1|^2 + \nonumber \\
& \phantom{= }\sum_{j\neq k} |(\h_k^{\herm}\Phib\G + \h_{\rmL,k}^{\herm})\w_j|^2 + \sigma_n^2 \\
& = \sum_j |(\h_k^{\herm}\Phib\G + \h_{\rmL,k}^{\herm})\w_j|^2 \nonumber \\
& \phantom{= }- 2 \, \Re\{(\h_k^{\herm}\Phib\G + \h_{\rmL,k}^{\herm})\w_k\} + 1 + \sigma_n^2.
\end{align}
\fi
The receive SMSE over all UEs is thus expressed as
\ifsingle
\begin{align}
\mathrm{SMSE} & = \sum_k \mathrm{MSE}_k \\
& = \sum_k \sum_j |(\h_k^{\herm}\Phib\G + \h_{\rmL,k}^{\herm})\w_j|^2 - 2 \sum_k \Re\{(\h_k^{\herm}\Phib\G + \h_{\rmL,k}^{\herm})\w_k\} + K(1 + \sigma_n^2). \label{eq:SMSE}
\end{align}
\else
\begin{align}
\mathrm{SMSE} & = \sum_k \mathrm{MSE}_k \\
& = \sum_k \sum_j |(\h_k^{\herm}\Phib\G + \h_{\rmL,k}^{\herm})\w_j|^2 \nonumber \\
& \phantom{= }- 2 \sum_k \Re\{(\h_k^{\herm}\Phib\G + \h_{\rmL,k}^{\herm})\w_k\} + K(1 + \sigma_n^2). \label{eq:SMSE}
\end{align}
\fi

For ease of presentation, let us define
\begin{align}\label{eq:v}
 \v = [\alpha_1e^{-j\phi_1},\ldots,\alpha_Ne^{-j\phi_N},1]^{\tran}\in \Compl^{N+1\times 1},   
\end{align}
and 
\begin{align}\label{eq:Hb}
\bar{\H}_k \triangleq \begin{bmatrix}
\mathrm{\diag}(\h_k^{\herm}) \G \\
\h_{\rmL,k}^{\herm}
\end{bmatrix}\in \Compl^{N+1\times M},
\end{align}
such that $\Phib = \mathrm{diag}(\v[1:N]^{\herm})$ and\,\footnote{Note that the last element of $\v$ is introduced to obtain a more compact expression of our optimization problems.}
\begin{align}
    (\h_{k}^{\herm}\Phib\G + \h_{\rmL,k}^{\herm})\w_j = \v^{\herm} \bar{\H}_k \w_j \quad \forall k,j.
\end{align}
Hence, our optimization problem can be formulated as the following
\begin{problem}[P\_SMSE]\label{eq:P_SMSE}
\begin{align}
   & \displaystyle \underset{\v ,\W}{\minimize} && \displaystyle \sum_k \sum_j |\v^{\herm}\bar{\H}_k\w_j|^2 - 2 \sum_k \Re\{\v^{\herm}\bar{\H}_k\w_k\} & \nonumber\\
   & \st && |v_i|^2 \leq 1, \quad i=1,\ldots,N; \nonumber \\
   &  && v_{N+1} = 1; \nonumber\\
   &  && \|\W\|^2_{\mathrm{F}} \leq P; \nonumber
\end{align}
\end{problem}
\noindent with $\v$ defined in Eq.~\eqref{eq:v} and $P$ the available transmit power at the BS. Note that the constraint $|v_i|^2\leq 1$ ensures that the $i$-th RIS element does not amplify the incoming signal, thus guaranteeing a passive structure overall. We remark that contrarily to previous works on beamforming optimization in RIS-aided networks \cite{Yu19, Nad19, Wu18, Mis19, Wu19, Kar19}, our proposed framework has the key advantage of being convex in the two optimization variables $\v$ and $\W$ separately. \emph{This allows us to find simple and efficient solutions to the problem at hand}. Moreover, thanks to this aforementioned key property the use of alternating optimization between the two optimization variables $\v$ and $\W$ allows us to guarantee convergence to a critical point of Problem~\ref{eq:P_SMSE}1, i.e., a point that satisfies the Karush-Kuhn-Tucker (KKT) conditions of Problem~\ref{eq:P_SMSE} (\!\cite{Gri00, Xu20}). Note that given the non convex nature of Problem~\ref{eq:P_SMSE}, the KKT conditions are necessary but not sufficient conditions for optimality. We now deeply examine our problem for two main use cases: $i$) single UE receiver and $ii$) multiuser receiver.

%

\section{Single user case}\label{sec:singleUE}

We firstly focus on the case of $K = 1$ to better highlight the key feature of the proposed \name{} method. In order to separately exploit the convexity in $\v$ and $\W$ of our objective function in Problem~\ref{eq:P_SMSE}, let the RIS parameters in $\v$ be fixed such that we can firstly focus on finding the precoding strategy $\W$. Since perfect CSI is available at the BS, when $\v$ is fixed the optimal linear transmit precoding vector is known to be the one matched to the (here, effective) channel between the BS and the UE maximizing the receive SNR, which is given by maximum-ratio transmission (MRT), i.e.,
\begin{align}\label{eq:mrt}
\w_{\mrt} = \sqrt{P} \frac{\G^{\herm}\Phib^{\herm}\h+ \h_{\rmL}}{\|\G^{\herm}\Phib^{\herm}\h+ \h_{\rmL}\|}.
\end{align}
Thus, once the precoding strategy is obtained the problem reduces to the optimization of the RIS setting parameters in $\Phib$. Consider the receive MSE after MRT precoding
\begin{align}
\mathrm{MSE}_{\mrt} = \Exp[|y-s|^2],
\end{align}
where the expectation is over the symbol $s$ and the noise $n$, which are assumed to be independent. Hence, we have that
\ifsingle
\begin{align}
\mathrm{MSE}_{\mrt} & = \Exp[|[(\h^{\herm}\Phib\G+ \h_{\rmL}^{\herm})\w_{\mrt}-1]s + n|^2] \\
& = P\|\h^{\herm}\Phib\G+ \h_{\rmL}^{\herm}\|^2 - 2 \sqrt{P}\|\h^{\herm}\Phib\G+ \h_{\rmL}^{\herm}\| + 1 + \sigma_n^2 \label{eq:mse}.
\end{align}
\else
\begin{align}
\mathrm{MSE}_{\mrt} & = \Exp[|[(\h^{\herm}\Phib\G+ \h_{\rmL}^{\herm})\w_{\mrt}-1]s + n|^2] \\
& = P\|\h^{\herm}\Phib\G+ \h_{\rmL}^{\herm}\|^2 - 2 \sqrt{P}\|\h^{\herm}\Phib\G+ \h_{\rmL}^{\herm}\| \nonumber
\\& \phantom{= } + 1 + \sigma_n^2 \label{eq:mse}.
\end{align}
\fi
We thus formulate the following optimization problem
\begin{align}\label{eq:P1}
\begin{array}{cl}
\displaystyle \underset{\Phib}{\minimize} & \displaystyle \|\h^{\herm}\Phib\G+ \h_{\rmL}^{\herm}\|^2 - 2\frac{\sqrt{P}}{P} \|\h^{\herm}\Phib\G+ \h_{\rmL}^{\herm}\| \\
\st & |[\Phib]_{ii}|^2 \leq 1 \quad \forall i \\
& [\Phib]_{ij} = 0 \quad \forall i\neq j.
\end{array}
\end{align}

By substituting Eq.~\eqref{eq:v} and Eq.~\eqref{eq:Hb} into Eq.~\eqref{eq:P1}, we recast the latter into the following optimization problem
\begin{problem}[P2]\label{eq:P2}
\begin{align}
   & \displaystyle \underset{\v}{\minimize} && \displaystyle \|\v^{\herm}\bar{\H}\|^2 - 2\frac{\sqrt{P}}{P}\|\v^{\herm}\bar{\H}\| & \nonumber\\
   & \st && |v_i|^2 \leq 1, \quad  i=1,\ldots,N; \nonumber\\
   &  && v_{N+1} = 1. \nonumber
\end{align}
\end{problem}
\noindent Note that Problem~\ref{eq:P2} is non-convex in $\v$ but it can be solved efficiently by standard convex-concave programming as it is a summation of a convex function, i.e., the squared norm term, minus a second convex function, i.e., the norm term \cite{She16}.

An alternative yet simpler approach defines $\V=\v\v^{\herm}$ and solve the following optimization problem
\begin{problem}[P3]\label{eq:P3}
\begin{align}
   & \displaystyle \underset{\v, \, \V\succeq \0}{\minimize} && \displaystyle \tr(\bar{\H}\bar{\H}^{\herm}\V) - 2 \frac{\sqrt{P}}{P}\sqrt{\tr(\bar{\H}\bar{\H}^{\herm}\V)} & \nonumber\\
   & \st && [\V]_{ii} \leq 1, \quad  i=1,\ldots,N; \nonumber\\
   &  && [\V]_{N+1,N+1} = 1, v_{N+1} = 1; \nonumber\\
   &  && \begin{bmatrix}
\V & \v \\
\v^{\herm} & 1
\end{bmatrix}\succeq \0, \nonumber\\
&  && \mathrm{rank}(\V) = 1. \nonumber
\end{align}
\end{problem}
%
\noindent Note that Problem~\ref{eq:P3} is non-convex in $\V$ due to the rank constraint. However, by employing SDR the latter can be turned into a convex problem by relaxing the rank constraint. The resulting problem can be then solved via standard semidefinite programming as, e.g., CVX. 
An approximate solution of Problem~\ref{eq:P3} can be obtained from the relaxed convex problem via Gaussian randomization~\cite{Luo10}. While the optimality of Gaussian Randomization is only proven for a small well-defined family of optimization problems, it guarantees an $\frac{\pi}{4}$-approximation of the optimal objective value of the original problem for a sufficiently large number of randomizations, as shown in~\cite{gaus_opt}.

Lastly, note that the RIS parameters $\{\alpha_i\}_{i=1}^N$ and $\{\phi_i\}_{i=1}^N$ can be obtained by setting
\begin{align}
& \alpha_i = |v_i|, \quad\text{and} \label{eq:alphai} \\ 
& \phi_i = \mathrm{arg}(v_i^*), \quad  i=1,\ldots,N. \label{eq:phii}
\end{align}

\subsection{Practical systems: low-resolution RIS}\label{subsec:Practical_systems_singleUE}
In practical systems, it is difficult to control exactly the state of each reflecting element as this control is implemented through sensible variations of the equivalent impedance of each reflecting cell. It is thus not practical to allow any possible state for the absorption coefficients $\{\alpha_i\}_{i=1}^N$ and phase reflection $\{\phi_i\}_{i=1}^N$ of the $i$-th reflecting element~\cite{Aru19, Hua19}. In this respect, we propose an extension of the method proposed in Section~\ref{sec:singleUE}, dubbed as Lo-\name{}, which decouples the optimization of $\{\alpha_i\}_{i=1}^N$ and $\{\phi_i\}_{i=1}^N$ to include practical implementation constraints, namely, each reflecting element is activated in a binary fashion and each phase shift can vary on a given set of discrete values.

{\bf Binary activation.}
\label{subsubsec:binary_singleUE}
We start by treating the binary activation assumption, namely each reflecting element can have only one of two states, i.e., $\alpha_i \in \{0,1\} \,\, \forall i$. Hence, we solve Problem~\eqref{eq:P2} or Eq.~\eqref{eq:P3} in order to obtain the values of $\{\phi\}_{i=1}^N$ as per Eq.~\eqref{eq:phii}. 

In the considered single-UE scenario, the maximization of the sum rate is equivalent to the minimization of the receive MSE or maximization of the receive SNR. Let us define the effective channel as the following
\begin{align}
\widetilde{\H} = \begin{bmatrix}
\mathrm{diag}(\h^{\herm}) \bar{\Phib} \G \\
\h_{\rmL}^{\herm}
\end{bmatrix}  \in \Compl^{N+1\times M},
\end{align}
with $\bar{\Phib} \triangleq \mathrm{diag}[e^{j\phi_1},\ldots,e^{j\phi_N}]$ and the binary vector $\b \in \{0,1\}^{N+1}$, where each $b_i$ indicates whether the corresponding reflecting element is active or not. Hence, we have that $\alpha_i= b_i \,\, i=1,\ldots,N$ and $\Phib = \mathrm{diag}(b_1 e^{j\phi_1},\ldots,b_N e^{j\phi_N})$. The receive SNR after MRT precoding is given by
\begin{align}
    \mathrm{SNR}_{\mrt} = \frac{\|\b^{\tran}\widetilde{\H}\|^2}{\sigma^2_n}, 
\end{align}
which is clearly maximed when $\b = \1$.

{\bf Quantized phase shifts.}
\label{subsubsec:quantized_phase_singleUE}
%
Consider now the case where the phases $\{\phi_i\}_{i=1}^N$ are quantized with a given number of bits $b$ as in explained in~\cite{Hua19, Wu19, Han19, Di20}. The ideal feasible set $[0,2\pi)$ is thus quantized into $2^b$ uniformly spaced discrete points as
\begin{align}\label{eq:phi_quantized}
    \phi_i \in \mathcal{Q} \triangleq \bigg\{\frac{2\pi}{2^b} \,m\bigg\}_{m=0}^{2^b-1} \quad m\in \mathbb{Z},\,\,i=1,\ldots,N.
\end{align}
To achieve such quantization, we simply project the phase shifts obtained by solving Problem~\ref{eq:P2} or \ref{eq:P3} onto the closest point within the constellation in $\mathcal{Q}$. 

\section{Multiuser case}\label{sec:multiUE}

Hereafter, we consider the multiuser scenario described in Section~\ref{sec:sys_model}. Differently than the single UE case, here the optimal transmit precoder is not know a priori and needs to be optimized. In particular, we show how the chosen optimization metric---which has not been analyzed so far in RIS-aided communication systems---yields simple expressions for both the optimized precoding strategy at the BS and the RIS parameters when employing alternating optimization between the two. Specifically, we solve the problem of jointly optimizing the precoding strategy and the RIS parameters by fixing in turn one of the two optimization variables and analyzing the resulting partial problems. It is interesting to see that such partial problems allow simple empirical closed-form solutions. The resulting algorithm provides an effective solution to the original joint optimization which can be proven to be a stationary point of the Lagrangian of the latter.

\subsection{Alternating Optimization}\label{subsec:AltOpt}

Let us consider Problem~\ref{eq:P_SMSE} (P\_SMSE), which is not jointly convex in $\v$ and $\W$ whereas, differently than prior work, is convex in the two optimization variables, separately. We can thus solve Problem~\ref{eq:P_SMSE} efficiently via alternating optimization. If $\W$ is fixed, then Problem~\ref{eq:P_SMSE} (P\_SMSE) reduces as follows
\begin{problem}[P\_SMSE\_v]\label{eq:P_SMSE_v}
\begin{align}
   & \displaystyle \underset{\v}{\minimize} && \displaystyle \sum_k\|\v^{\herm} \bar{\H}_k \W\|^2 - 2 \sum_k \Re\{\v^{\herm}\bar{\H}_k \w_k\} & \nonumber\\
   & \st && |v_i|^2 \leq 1 \quad i=1,\ldots,N; \nonumber\\
   &  && v_{N+1} = 1. \nonumber
\end{align}
\end{problem}
\noindent Problem~\ref{eq:P_SMSE_v} admits the following solution
\ifsingle
\begin{align}\label{eq:v_opt_AO}
   \v & = \bigg(\sum_k \bar{\H}_k\W\W^{\herm}\bar{\H}^{\herm}_k + \mathrm{Diag}(\mub)\bigg)^{-1}  \bigg(\sum_k \bar{\H}_k \w_k - \nu \, \e_{N+1}\bigg)
\end{align}
\else
\begin{align}\label{eq:v_opt_AO}
   \v & = \bigg(\sum_k \bar{\H}_k\W\W^{\herm}\bar{\H}^{\herm}_k + \mathrm{Diag}(\mub)\bigg)^{-1} \nonumber \\
   & \times \bigg(\sum_k \bar{\H}_k \w_k - \nu \, \e_{N+1}\bigg),
\end{align}
\fi
where $\e_{N+1}$ is the $(N+1)$-th column of the identity matrix of size $N+1$ and $\mub \geq \0$ is a vector of non-negative variables to be determined in the following way
\begin{align}
    & \mu_i = 0 \text{~and~} |v_i|^2 \leq 1, \nonumber \\
    & \mu_i \geq 0 \text{~and~} |v_i|^2 = 1, \quad \forall i=1,\ldots,N. \label{eq:mub}
\end{align}
\noindent To alleviate the task of finding $\mub$ in Eq.~\eqref{eq:mub} we set $\mub = \sigma_n^2 \1$ following the results in \cite{Pee05}. Let
\begin{align}
    \bar{\v} & = \bigg(\sum_k \bar{\H}_k\W\W^{\herm}\bar{\H}^{\herm}_k + \sigma_n^2 \I_{N+1}\bigg)^{-1} \nonumber \\
   & \times \bigg(\sum_k \bar{\H}_k \w_k - \nu \, \e_{N+1}\bigg)
\end{align}
where $\I_{N+1}$ is the identity matrix of size $N+1$. Hence we have that
\begin{align}
    \v & = \frac{\bar{\v}}{\|\bar{\v}\|}.
\end{align}
Lastly, $\nu$ is found by letting $v_{N+1} = 1$ as
\begin{align}\label{eq:nu}
    \nu = \frac{\e_{N+1}^{\tran} \B \, \z - 1}{\e_{N+1}^{\tran}\B \,\e_{N+1}}.
\end{align}
where
\begin{align}
    \B = \bigg(\sum_k \bar{\H}_k\W\W^{\herm}\bar{\H}^{\herm}_k + \sigma_n^2 \I_{N+1}\bigg)^{-1},
\end{align}
$\z = \sum_k \bar{\H}_k \w_k$ and $\e_{N+1}$ is the $(N+1)$-th column of the identity matrix of size $N+1$.
\begin{IEEEproof}
The solution of Problem~\ref{eq:P_SMSE_v} is analytically derived in Appendix~\ref{ap:SMSE_v} by solving the Karush–Kuhn–Tucker (KKT) conditions.
\end{IEEEproof}

When $\v$ is fixed, Problem~\ref{eq:P_SMSE}(P\_SMSE) reduces to the following
\begin{problem}[P\_SMSE\_w]\label{eq:P_SMSE_w}
\begin{align}
   & \displaystyle \underset{\W}{\minimize} && \displaystyle \sum_k\|\bar{\h}^{\herm}_k\W\|^2 - 2 \sum_k \Re\{\bar{\h}^{\herm}_k\W \e_k\} & \nonumber\\
   & \st && \|\W\|^2_{\mathrm{F}} \leq P; \nonumber
\end{align}
\end{problem}
\noindent where we define $\bar{\h}_k \triangleq \bar{\H}_k^{\herm}\v$ and $\e_k$ is the $k$-th column of the identity matrix of size $K$. Again, given the convexity of Problem~\ref{eq:P_SMSE_w}, the KKT conditions are necessary and sufficient for the solution of the problem and yield the following 
\begin{align}
    \W = \big(\bar{\H}\bar{\H}^{\herm} + \mu\I_M\big)^{-1}\bar{\H} \label{eq:W_opt_AO},
\end{align}
with $\bar{\H} = [\bar{\h}_1,\ldots,\bar{\h}_K]$ and $\mu\geq 0$ such that $\|\W\|_{\mathrm{F}}^2=P$ is satisfied\footnote{In order to determine $\mu$ we can apply a bisection method.}. Leveraging the results in \cite{Pee05} we set $\mu = K\sigma_n^2/P$ which is proven to maximize the UEs SINRs in the limit of large $K$, while proving to be tight for even small values of $K$. Hence we obtain the following empirical closed-form expression for the precoding matrix $\W$
\begin{align}
    \W = \sqrt{P} \frac{\bar{\W}}{\|\bar{\W}\|_{\mathrm{F}}}
\end{align}
where $\bar{\W} = \big(\bar{\H}\bar{\H}^{\herm} + \frac{K\sigma_n^2}{P}\I_M\big)^{-1}\bar{\H}$.
\begin{IEEEproof}
We derive Eq.~\eqref{eq:W_opt_AO} in Appendix \ref{ap:SMSE_w} by solving the KKT conditions.
\end{IEEEproof}

Due to the convex nature of Problem~\ref{eq:P_SMSE_v} and \ref{eq:P_SMSE_w}, we propose an efficient algorithm, namely \name{}, which alternates the optimization of both the precoding strategy at the BS $\W$ and the RIS setting parameters in $\v$. Thanks to the convex nature of the two partial problems for which we have found an optimal solution, it can be proven that \name{} converges to a critical point of Problem~\ref{eq:P_SMSE}, i.e., a point that satisfies the KKT conditions of Problem~\ref{eq:P_SMSE} (\!\cite{Gri00,Xu20}). The proposed algorithm is formally described in Algorithm~\ref{alg:A1_2} where step \ref{alg:A1_2_statenu} implements Eq.~\eqref{eq:nu}. 

\begin{algorithm}[t!]
  \caption{\name: RIS-aided Multiuser Alternating optimization}\label{alg:A1_2}
  \begin{algorithmic}[1]
     \State Initialize $\W^{(0)}$, $\mathrm{SMSE}^{(0)}$ and $\epsilon$ 
      \State $n\gets 1$
     \While { $|\big(\mathrm{SMSE}^{(n)} - \mathrm{SMSE}^{(n-1)}\big)/\mathrm{SMSE}^{(n)}| > \epsilon$ }
     \State $\mub = \sigma_n^2 \1$ \label{alg:A1_2_statemub}
     \State $\nu \gets \text{Force }v_{N+1}^{(n)}=1$\label{alg:A1_2_statenu}
     \ifsingle
     \State $$\bar{\v} = \bigg(\sum_k \bar{\H}_k\W\W^{\herm}\bar{\H}^{\herm}_k + \mathrm{Diag}(\mub)\bigg)^{-1} \bigg(\sum_k \bar{\H}_k \w_k - \nu \, \e_{N+1}\bigg)$$
     \else
     \State $$\bar{\v} = \bigg(\sum_k \bar{\H}_k\W\W^{\herm}\bar{\H}^{\herm}_k + \mathrm{Diag}(\mub)\bigg)^{-1}$$ \linebreak[0] $$\times \bigg(\sum_k \bar{\H}_k \w_k - \nu \, \e_{N+1}\bigg)$$
     \fi
     \State  $$\v^{(n)} =  \frac{\bar{\v}}{\|\bar{\v}\|}$$
     \State $\bar{\H} = [\bar{\H}_1^{\herm}\v^{(n)},\ldots,\bar{\H}_K^{\herm}\v^{(n)}]$
     \State $$\mu = \frac{K\sigma_n^2}{P}$$
     \State $\bar{\W} = \big(\bar{\H}\bar{\H}^ {\herm} + \mu\I_M\big)^{-1}\bar{\H}$
     \State $$\W^{(n)} = \sqrt{P} \frac{\bar{\W}}{\|\bar{\W}\|_{\mathrm{F}}}$$
     \ifsingle
     \State $$\mathrm{SMSE}^{(n)} = \sum_k \sum_j |(\v^{(n)})^{\herm}\bar{\H}_k\w_j^{(n)}|^2- 2 \sum_k \Re\{(\v^{(n)})^{\herm}\bar{\H}_k\w_k^{(n)}\} + K(1 + \sigma_n^2)$$
     \else
     \State $$\mathrm{SMSE}^{(n)} = \sum_k \sum_j |(\v^{(n)})^{\herm}\bar{\H}_k\w_j^{(n)}|^2$$ \linebreak[0] $$- 2 \sum_k \Re\{(\v^{(n)})^{\herm}\bar{\H}_k\w_k^{(n)}\} + K(1 + \sigma_n^2)$$
     \fi
     \EndWhile
     \State $\v = \v^{(n)}$
     \State $\W =\W^{(n)}$
  \end{algorithmic}
\end{algorithm}
\subsection{Practical systems: low resolution RIS}\label{subsec:Practical_systems_multiUE}
As described in Section~\ref{subsec:Practical_systems_singleUE}, in practical conditions it is difficult to control the state of each reflecting element perfectly. In the following, we reformulate the problem in Section~\ref{subsec:AltOpt} to cope with the limits of practical hardware implementations and we assume that each reflecting element can be activated in a binary fashion and introduces only quantized phase shifts. The proposed algorithm dubbed Lo-\name{} is formally described in Algorithm~\ref{alg:A1_bin_2}.

\begin{algorithm}[t!]
  \caption{Lo-\name{}: Low-Resolution RISMA Algorithm}\label{alg:A1_bin_2}
  \begin{algorithmic}[1]
     \State Initialize $\W^{(0)}$, $\mathrm{SMSE}^{(0)}$ and $\epsilon$ 
      \State $n\gets 1$
     \While { $|\big(\mathrm{SMSE}^{(n)} - \mathrm{SMSE}^{(n-1)}\big)/\mathrm{SMSE}^{(n)}| > \epsilon$ }
     \State $\widetilde{\H}_k =  \begin{bmatrix}
    \bar{\H}_k\W\W^{\herm}\bar{\H}_k^{\herm} & -\bar{\H}_k\w_k \\
    -\w_k^{\herm}\bar{\H}_k^{\herm} & 0
    \end{bmatrix}, \quad k=1,\ldots, K$
     \State $$\hat{\v}^{(n)} = \argmin_{\bar{\v}} \sum_k \bar{\v}^{\herm}\widetilde{\H}_k\bar{\v} $$ \linebreak[0] $$\qquad \qquad \qquad \qquad \qquad \qquad \quad \mathrm{s.t.} \quad \bar{\v} \in \bar{\mathcal{Q}}^{N+1}, \quad \bar{\v} = [\v \ c], \quad |c|^2=1$$
     \State $\v^{(n)} = (c^{(n)})^*\hat{\v}^{(n)}$
     \State $\bar{\H} = [\bar{\H}_1^{\herm}\v^{(n)},\ldots,\bar{\H}_K^{\herm}\v^{(n)}]$
     \State $$\mu = \frac{K\sigma_n^2}{P}$$
     \State $\bar{\W} = \big(\bar{\H}\bar{\H}^ {\herm} + \mu\I_M\big)^{-1}\bar{\H}$
     \State $$\W^{(n)} = \sqrt{P} \frac{\bar{\W}}{\|\bar{\W}\|_{\mathrm{F}}}$$
     \ifsingle
     \State $$\mathrm{SMSE}^{(n)} = \sum_k \sum_j |(\v^{(n)})^{\herm}\bar{\H}_k\w_j^{(n)}|^2- 2 \sum_k \Re\{(\v^{(n)})^{\herm}\bar{\H}_k\w_k^{(n)}\} + K(1 + \sigma_n^2)$$
     \else
     \State $$\mathrm{SMSE}^{(n)} = \sum_k \sum_j |(\v^{(n)})^{\herm}\bar{\H}_k\w_j^{(n)}|^2$$ \linebreak[0] $$- 2 \sum_k \Re\{(\v^{(n)})^{\herm}\bar{\H}_k\w_k^{(n)}\} + K(1 + \sigma_n^2)$$
     \fi
     \EndWhile
     \State $\v = \v^{(n)}$
     \State $\W =\W^{(n)}$
  \end{algorithmic}
\end{algorithm}
If $\W$ is fixed then Problem~\ref{eq:P_SMSE_v} (P\_SMSE\_v) stated in Section~\ref{subsec:AltOpt} is modified as follows
\begin{problem}[P\_SMSE\_Lo]\label{eq:P_SMSE_Lo}
\begin{align}
   & \displaystyle \underset{\v}{\minimize} && \displaystyle \sum_k\|\v^{\herm} \bar{\H}_k \W\|^2 - 2 \sum_k \Re\{\v^{\herm}\bar{\H}_k \w_k\} & \nonumber\\
   & \st && v_i \in \bar{\mathcal{Q}} \quad i=1,\ldots,N; \nonumber\\
   &  && v_{N+1} = 1, \nonumber
\end{align}
\end{problem}
\noindent where we have defined the constellation of discrete points $\bar{\mathcal{Q}}$ as
\begin{align}\label{eq:Q_bar}
    \bar{\mathcal{Q}} \triangleq \bigg\{0, \ e^{j \frac{2\pi}{2^b} \,m}\bigg\}_{m=0}^{2^b-1} \quad m\in \mathbb{Z},\,\,i=1,\ldots,N;
\end{align}
to include the deactivated RIS antenna elements and the quantized phase shifts. Let the effective channel matrix of the $k$-th UE be defined as follows
\begin{align}
    \widetilde{\H}_k = \begin{bmatrix}
    \bar{\H}_k\W\W^{\herm}\bar{\H}_k^{\herm} & -\bar{\H}_k\w_k \\
    -\w_k^{\herm}\bar{\H}_k^{\herm} & 0
    \end{bmatrix} \in \Compl^{N+2\times N+2}.
\end{align}
Lastly, let us define $\bar{\v} \triangleq [\v^{\tran} \ c]^{\tran}$ and $\bar{\V} = \bar{\v}\bar{\v}^{\tran}$ with $|c|^2 = 1$. Hence, Problem~\eqref{eq:P_SMSE_Lo} (P\_SMSE\_Lo) is equivalent to the following homogeneous quadratic problem
\begin{problem}[P\_SMSE\_LoE]\label{eq:P_SMSE_LoE}
\begin{align}
   & \displaystyle \underset{\bar{\v},\bar{\V}}{\minimize} && \displaystyle \sum_k\tr(\widetilde{\H}_k \bar{\V})  & \nonumber\\
   & \st && \mathrm{diag}(\bar{\V}) \in \{0,1\}; \nonumber\\
   &  && \bar{\V}_{N+1,N+1} = \bar{\V}_{N+2,N+2} = 1; \nonumber \\
   & && \begin{bmatrix}\bar{\V} & \bar{\v} \\ \bar{\v}^{\tran} & 1 \end{bmatrix}\succeq \0; \nonumber\\
   & && \mathrm{rank}(\bar{\V}) = 1. \nonumber
\end{align}
\end{problem}
Following the results in \cite{Ma04, Sid06}, we relax Problem~\ref{eq:P_SMSE_LoE} (P\_SMSE\_LoE) by removing the rank constraint and substituting the binary constraint on the diagonal of $\bar{\V}$ with the convex constraint $\0 \leq \mathrm{diag}(\bar{\V})\leq \1$. Let $\bar{\V}^*$ denote the solution of Problem~\ref{eq:P_SMSE_LoE}. By applying the Gaussian randomization method in \cite{Luo10}, we use $\bar{\V}^*$ to generate $L$ random vectors $\w_l \sim \mathcal{CN}(\0,\bar{\V}^*)$. Such vectors are then quantized into the nearest point within the constellation $\bar{\mathcal{Q}}$, thus obtaining the set of $L$ vectors $\{\bar{\w}_l\}$. We then obtain an approximate solution of Problem~\ref{eq:P_SMSE_LoE} as follows
\begin{align}
\bar{\w}^{\star} =  \argmin_{l=1,\ldots,L} \sum_k \bar{\w}_l^{\tran}\widetilde{\H}_k\bar{\w}_l.
\end{align}
Let us obtain $\hat{\w}^{\star}$ from $\bar{\w}^{\star} = [(\hat{\w}^{\star})^{\herm} \ c^{\star}]^\herm$. A suboptimal solution to Problem~\ref{eq:P_SMSE_Lo} (P\_SMSE\_Lo) is thus given by
\begin{align}
    \v = c^{\star}\hat{\w}^{\star}.
\end{align}
Lastly, note that such optimization framework can be readily extended to any discrete set of phase shifts by simply changing the definition of $\bar{\mathcal{Q}}$ in Eq.~\eqref{eq:Q_bar} and modifying accordingly the quantization operation on the random vectors $\{\w_l\}$.

\section{Numerical Results and Discussion}\label{sec:Results}
We present numerical results to analyze the benefits of the proposed algorithms both in the single UE case as per Section~\ref{sec:singleUE} and in a general multiuser setting as per Section~\ref{sec:multiUE}. Moreover, we show that our proposed scheme provides substantial gains compared to conventional massive MIMO schemes such as zero-forcing (ZF) or minimum mean squared error (MMSE) precoding (without the aid of RIS) as the transmit power $P$, the number of BS antennas $M$ and the network area radius vary on a broad range of values. Finally, we show that similar outstanding gains can be attained even when considering RIS as a low-resolution surface whose antenna elements are activated in a binary fashion thereby introducing only uniformly spaced discrete phase shifts. 

\ifsingle
\begin{figure}[t!]
    \centering
    \includegraphics[scale=0.25]{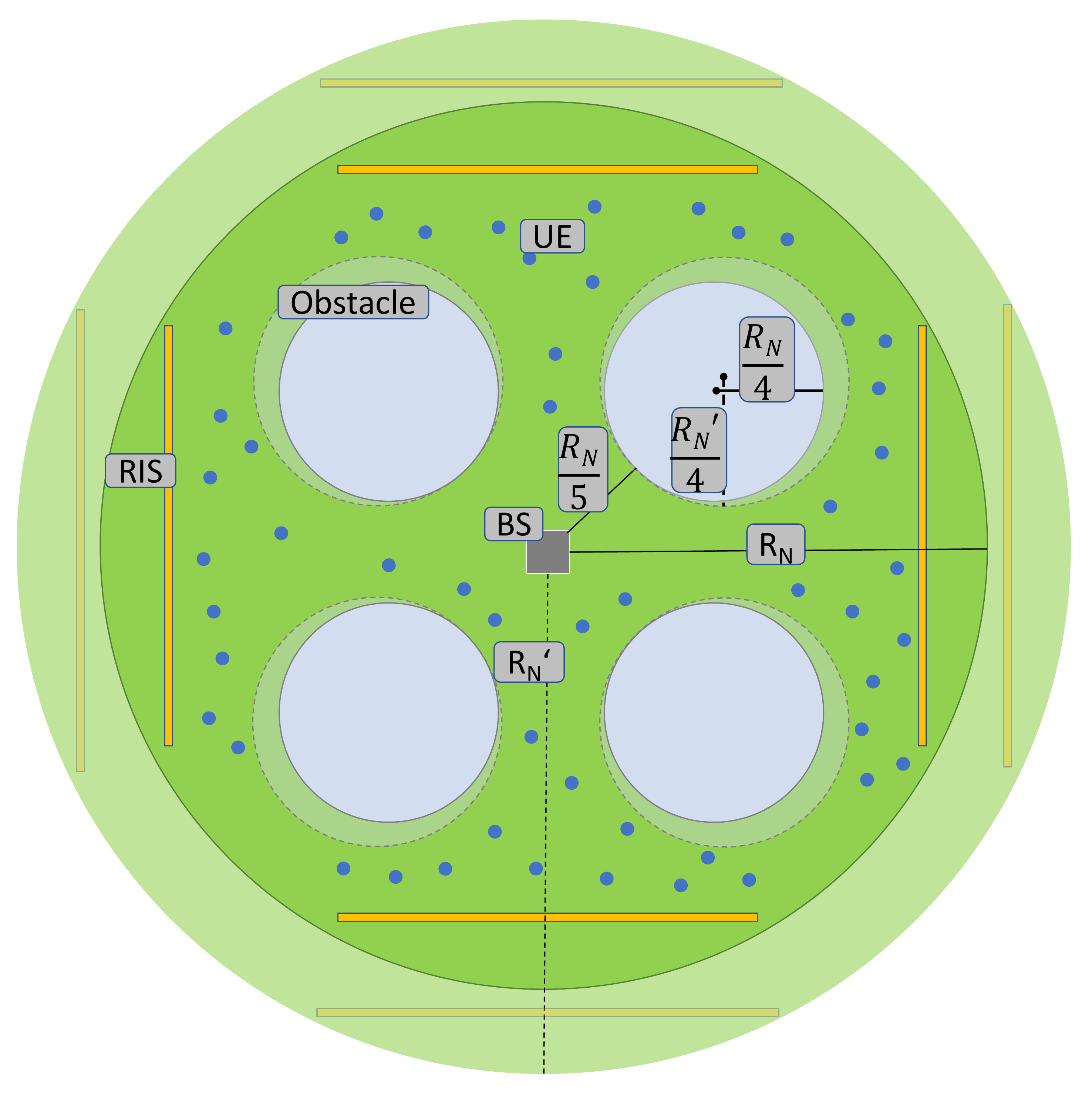}\vspace{-4mm}
    \caption{Single-BS scenario where multiple obstacles and RISs are placed while UEs are normally distributed within a $R_{\text{N}}$-radius area.}
    \label{fig:eval_scenario}
\end{figure}
\else
\begin{figure}[t!]
    \centering
    \includegraphics[width=0.8\columnwidth, clip]{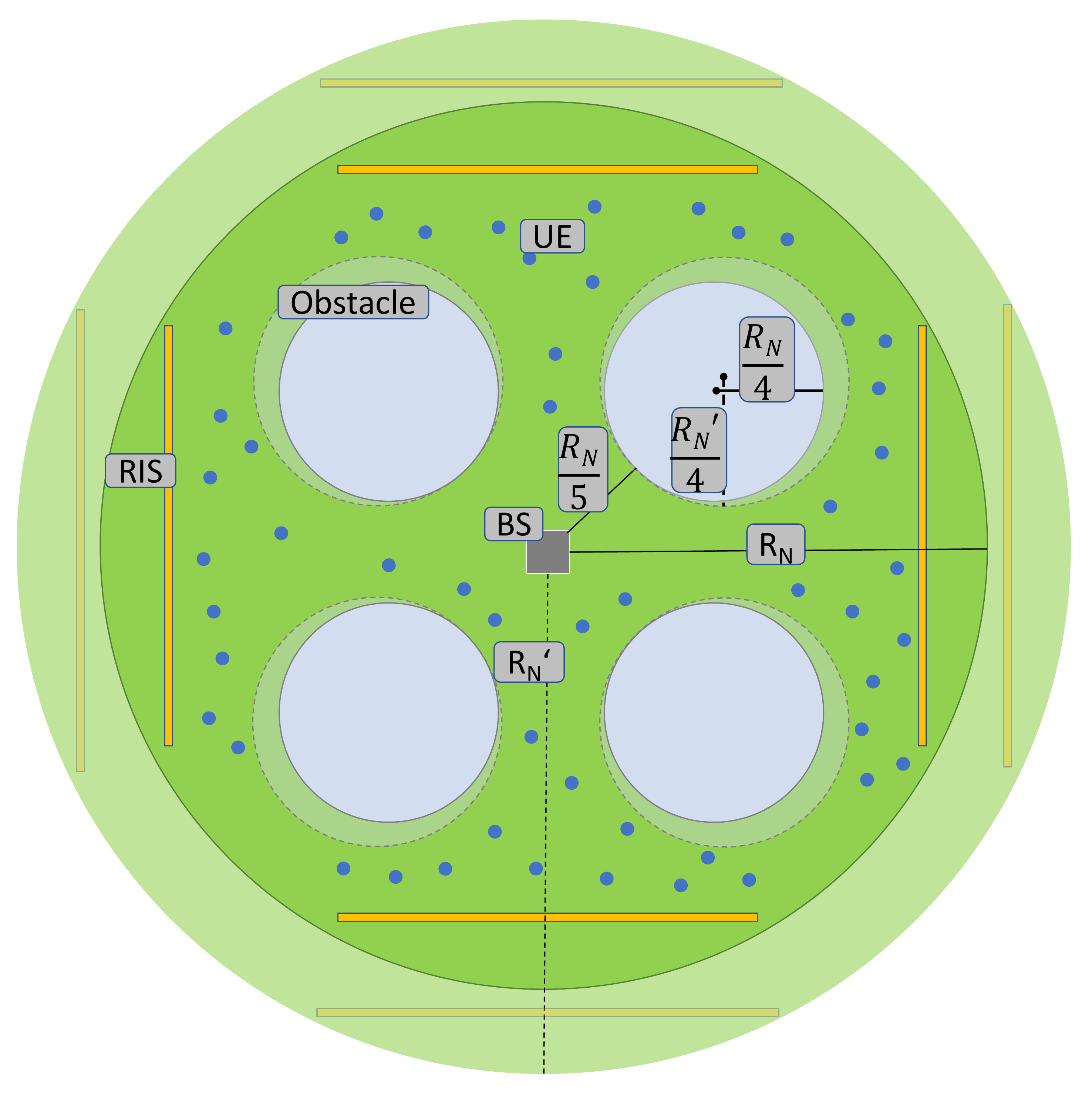}\vspace{-4mm}
    \caption{Single-BS scenario where multiple obstacles and RISs are placed while UEs are normally distributed within a $R_{\text{N}}$-radius area.}
    \label{fig:eval_scenario}
\end{figure}
\fi

\subsection{Channel model}\label{subsec:channel_model}
The channel model used for our numerical results is defined as follows. Let $\h_{\rmL}$ denote the direct channel between the BS and UE~$k$ defined as follows
\begin{align}
    \h_{\rmL,k}\!\triangleq\!\sqrt{\!\frac{K_{\rmR,\rmL,k}}{K_{\rmR,\rmL,k}+1}} \h_{\rmL,k}^{\Los}\!+\!\sqrt{\frac{1}{1+K_{\rmR,\rmL,k}}} \h_{\rmL,k}^{\Nlos}\!\!\in\! \Compl^{M\!\times\! 1}\!,
\end{align}
where $K_{\rmR,\rmL,k}$ is the Rician factor of UE~$k$ associated with the direct path from the BS to UE~$k$ while $\h_{\rmL,k}^{\Los}$ and $\h_{\rmL,k}^{\Nlos}$ denote LoS and NLoS components, respectively. The former is defined as
\begin{align}
    \h_{\rmL,k}^{\Los} \triangleq \sqrt{\gamma_{\rmL,k}} \, \a(\theta_k) \in \Compl^{M\times 1},
\end{align}
where $\gamma_{\rmL,k} = d_k^{-\beta_{\rmL,k}}$ is the large-scale fading coefficient with $\beta_{\rmL,k}$ the pathloss exponent and $\a(\theta_k)$ is the uniform linear array (ULA) response vector at the BS for the given steering angle $\theta_k$ defined as
\begin{align}
    \a(\theta_k) \triangleq [1,e^{j2\pi\delta \cos(\theta_k)},\ldots,e^{j2\pi\delta(M-1)\cos(\theta_k)}]^{\tran} \in \Compl^{M\times 1}.
\end{align}
Here, $\delta$ is the antenna spacing-wavelength ratio. Similarly, we define the NLoS component as the following
\begin{align}
    \h_{\rmL,k}^{\Nlos} \triangleq \sqrt{\frac{\gamma_{\rmL,k}}{P_{\rmL,k}}} \sum_{p=1}^{P_{\rmL,k}} \eta_{\rmL,k,p}  \, \a(\theta_{k,p}) \in \Compl^{M\times 1},
\end{align}
where $\eta_{\rmL,k,p}\sim \mathcal{CN}(0,1)$, $P_{\rmL,k}$ and $\theta_{k,p}$ are the small-scale fading coefficient, the number of scattered paths and the steering angle of the $p$-th scattered path between the BS and UE~$k$, respectively. We denote the channel between the BS and the RIS as follows
\begin{align}
    \G & \triangleq \sqrt{\frac{K_{\rmR}}{1+K_{\rmR}}}
    \, \G^{\Los} + \sqrt{\frac{1}{1+K_{\rmR}}}\, \G^{\Nlos} \in \Compl^{N\times M},
\end{align}
where $K_{\rmR}$ is the Rician factor, whereas $\G^{\Los}$ and $\G^{\Nlos}$ represent the deterministic LoS and Rayleigh fading components, respectively. The latter component is defined as
\begin{align}
    \G^{\Los}\triangleq \sqrt{\gamma_G} \, \b(\psi_{A})\a(\psi_{D})^{\herm} \in \Compl^{N\times M},
\end{align}
where $\gamma_G = d_{1}^{-\beta}$ is the large-scale fading coefficient with $\beta$ the pathloss exponent and $\b(\psi_{A})$ is the planar linear array (PLA) response vector, which models the RIS response for the steering angle $\psi_{A}$. We assume that the RIS is a two dimensional structure with $N=N_x N_y$ elements where $N_x$ and $N_y$ are the number of elements along the $x$ and $y$ axis, respectively. The PLA response is defined as
\ifsingle
\begin{align}
    \b(\psi_{A}) & \triangleq \b_z(\psi_{A,z},\psi_{A,x}) \otimes \b_x(\psi_{A,z},\psi_{A,x}) \\
    & = [1,e^{-j2\pi\delta \sin(\psi_{A,z})\cos(\psi_{A,x})},\ldots,  e^{-j2\pi\delta (N_y-1)\sin(\psi_{A,z})\cos(\psi_{A,x})}]^{\tran} \nonumber \\ & \otimes   [1,e^{-j2\pi\delta \cos(\psi_{A,x})\cos(\psi_{A,z})},\ldots,e^{-j2\pi\delta (N_x-1)\cos(\psi_{A,x})\cos(\psi_{A,z})}]^{\tran},
\end{align}
\else
\begin{align}
    \b(\psi_{A}) & \triangleq \b_z(\psi_{A,z},\psi_{A,x}) \otimes \b_x(\psi_{A,z},\psi_{A,x}) \\
    & = [1,e^{-j2\pi\delta \sin(\psi_{A,z})\cos(\psi_{A,x})},\ldots, \nonumber \\ & \phantom{= } e^{-j2\pi\delta (N_y-1)\sin(\psi_{A,z})\cos(\psi_{A,x})}]^{\tran} \otimes  \nonumber \\ & \phantom{= } [1,e^{-j2\pi\delta \cos(\psi_{A,x})\cos(\psi_{A,z})},\ldots,\nonumber \\ & \phantom{= }e^{-j2\pi\delta (N_x-1)\cos(\psi_{A,x})\cos(\psi_{A,z})}]^{\tran},
\end{align}
\fi
with $\psi_{A,z}$ and $\psi_{A,x}$ the azimuth and longitudinal AoA, repsectively. The NLoS component of the BS-RIS link is defined as
\begin{align}
    \G^{\Nlos}\!\!\triangleq\!\!\sqrt{\frac{\gamma_G}{P_G}} \!\sum_{p=1}^{P_G} \!\G^{(w)}_p\! \circ\! \big(\b(\psi_{A,p})\a(\psi_{D,p})^{\herm}\big)\!\in\! \Compl^{N\times M}
\end{align}
where $P_G$ is the total number of scattered paths, $\G^{(w)}$ represents the small-scale fading coefficients of the $p$-th path with $\mathrm{vec}(\G^{(w)}_p)\sim \mathcal{CN}(\0,\I_{NM})$, $\circ$ stands for element-wise product and $\psi_{A,p}$ and $\psi_{D,p}$ are the AoA and AoD of the $p$-th path, respectively. Lastly, $\h_k$ denotes the channel between the RIS and UE~$k$ defined as follows
\begin{align}
    \h_k\! \triangleq\! \sqrt{\frac{K_{\rmR,k}}{K_{\rmR,k}+1}} \h_{k}^{\Los}\! +\! \sqrt{\frac{1}{1+K_{\rmR,k}}} \h_{k}^{\Nlos}\! \in\! \Compl^{N\times 1},
\end{align}
where $K_{\rmR,k}$ is the Rician factor of UE~$k$ while $\h_{k}^{\Los}$ and $\h_{k}^{\Nlos}$ denote the LoS and NLoS components, respectively. The LoS component of the RIS-UE~$k$ link is defined as
\begin{align}
    \h_{k}^{\Los} =  \sqrt{\gamma_k} \, \b(\psi_k) \in \Compl^{N\times 1},
\end{align}
where $\gamma_k = d_{2,k}^{-\beta_k}$ is the large-scale fading coefficient and $\beta_k$ the pathloss exponent. The NLoS component of the RIS-UE~$k$ link is defined as the following
\begin{align}
    \h_{k}^{\Nlos} \triangleq \sqrt{\frac{\gamma_{k}}{P_{k}}} \sum_{p=1}^{P_{k}} \eta_{k,p}  \, \b(\psi_{k,p}) \in \Compl^{N\times 1},
\end{align}
where $\eta_{k,p}\sim \mathcal{CN}(0,1)$, $P_{k}$ and $\psi_{k,p}$ denote the small-scale fading coefficient, the number of scattered paths and the steering angle of the $p$-th scattered path related to the RIS-UE~$k$ link, respectively. 

\subsection{Power scaling law}\label{subsec:power_scaling}
We derive the power scaling law of the channel model proposed in Section~\ref{subsec:channel_model}. For the sake of clarity, we focus on a single UE and single BS antenna case, i.e., $M=1$ and hence $G \equiv g$ and $\w_{\mrt} \equiv w_{\mrt}$. Moreover, we assume that $\h \sim \mathcal{CN}(\0,\gamma \I_N)$, $\g \sim \mathcal{CN}(\0,\gamma_G \I_N)$ and $h_{\rmL} \sim \mathcal{CN}(0,\gamma_{\rmL})$. The average receive power at the UE is given by
\begin{align}
    P_{\rmUE} & = \Exp [|(\h^{\herm}\Phib\g + h_{\rmL}^{\herm})w_{\mrt}|^2] \label{eq:P_UE_indep} \\ 
    & = P (\Exp[|\h^{\herm}\Phib\g|^2] + \Exp[|h_{\rmL}|^2]) \nonumber\\
    & = P \bigg( \Exp\bigg[\bigg|\sum_{i=1}^N |h_i| |g_i| e^{j(\phi_i - \mathrm{arg}(h_i) + \mathrm{arg}(g_i))}\bigg|^2\bigg] + \gamma_{\rmL}\bigg) \nonumber\\ 
    & \leq P \left(\frac{\pi^2}{16}d_2^{-\beta_1} \, d_1^{-\beta} \, N^2+ d^{-\beta_{\rmL}}\right).\label{eq:P_UE_max}
    \end{align}
In Eq.~\eqref{eq:P_UE_indep} we assume that $\h$, $\g$ and $h_{\rmL}$ are statistically independent and Eq.~\eqref{eq:P_UE_max} follows by assuming optimal choice of the RIS phase shifts as $\phi_i = \mathrm{arg}(h_i) - \mathrm{arg}(g_i), \, \forall i$ and the fact that $\Exp[|h_i|^2|g_i|^2] = \pi^2 \gamma \, \gamma_G / 16$. The receive power thus scales as the inverse of the product of the distance of the individual paths from the BS to the RIS and from the RIS to the UE. Additionally, it scales as the square of the number of the RIS reflecting elements $N$ (in accordance with recent works on pathloss modelling~\cite{Ozd19, Tan19, Bjo19, bjrnson2019demystifying}). Hence, by increasing the number of RIS antenna elements, we can counteract the decrease in receive power due to the distance of the combined path from the BS to RIS and from the RIS to the UE. This notably suggests that \emph{RISs can be used smartly to effectively increase the coverage area of wireless networks}. 

\subsection{Scenario and setting parameters}\label{subsec:params}
We consider a circular single-cell network of radius $R_N$ with a central BS as depicted in Fig.~\ref{fig:eval_scenario}. Here, our goal is to evaluate the performance of the proposed scheme in terms of the sum rate defined in Eq.~\eqref{eq:srate} while increasing the radius of the network area to prove how the considered RIS-aided network is effective in increasing the coverage area of cellular networks. Indeed, we vary $R_N$ for a fixed number of UEs $K$. 

For each value of $R_N$ we average our simulations over $1000$ different realizations of the UEs' positions, according to a uniform distribution over the considered circular area. There are four circular obstacles, which determine whether each UE is in LoS with the BS or not. To maintain consistency, we fix the radii of such obstacles to $R_N/4$ and the distance from the BS to the centers of said obstacles to $R_N/4 \, + R_N/5$. Moreover, we assume that there are four different RISs in LoS with the BS, each one with $N_x = N_y = \sqrt{N}$ antenna elements, positioned at distance $R_N$ from the BS and angles $0$, $\pi/2$, $\pi$ and $3\pi/2$, respectively. Hence, for each RIS $d_{1} = R_N$, $\psi_{A,x} = \pi$, $3\pi/2$, $0$ and $\pi/2$ with $\psi_{A,z} = 0$, respectively. Furthermore, the AoD are $\psi_D = 0$, $\pi/2$, $\pi$ and $3\pi/2$, respectively. Each UE $k$ is served by a single RIS, according to the highest average channel power gain $\delta_k$ of the corresponding link, defined as the following
\begin{align}
    \delta_k & = \Exp[\|\h_k\|^2]\nonumber\\
    & = \frac{K_{\rmR,k}}{1+K_{\rmR,k}} \gamma_k N + \frac{1}{1+K_{\rmR,k}} \gamma_k N\nonumber \\
    & = d_{2,k}^{-\beta_k} \, N.
\end{align}
We assume that $\beta_k = \bar{\beta}, \, \forall k$ such that each UE $k$ is served by the closest RIS in terms of distance\,\footnote{Note that when (unlikely) more than one RIS is at the same distance from UE $k$, we solve the conflict by simply flipping a coin.}. For simplicity we let $K_{\rmR,k} = \bar{K}_{\rmR}$, $P_k = \bar{P}$, $P_{\rmL,k} = P_{\rmL}$, $K_{\rmL,k} = K_{\rmL}$ and $\beta_{\rmL,k} = \beta_{\rmL},\, \forall k$. We set $\bar{\beta}=\bar{\beta}_{\Los}$, $\bar{K}_{\rmR} = \bar{K}_{\rmR,\Los}$, $K_{\rmL} = K_{\rmL,\Los}$ and $\beta_{\rmL} = \beta_{\rmL,\Los}$ for LoS UEs while we let $\bar{\beta} = \bar{\beta}_{\Nlos}$, $\bar{K}_{\rmR} = \bar{K}_{\rmR,\Nlos}$, $K_{\rmL} = K_{\rmL,\Nlos}$ and $\beta_{\rmL} = \beta_{\rmL,\Nlos}$ for NLoS UEs. All simulations parameters are set as per Table \ref{tab:channel_param}, unless otherwise stated.

\begin{table}[h!]
\caption{Simulation parameters}
\label{tab:channel_param}
\vspace{-3mm}
\centering
\ifsingle
\resizebox{0.7\textwidth}{!}{%
\renewcommand{\arraystretch}{1.2}
\begin{tabular}{c|c|c|c|c|c|c|c}

\cellcolor[HTML]{EFEFEF} \textbf{Param.} & \textbf{Val.} & \cellcolor[HTML]{EFEFEF} \textbf{Param.}& \textbf{Val.} & \cellcolor[HTML]{EFEFEF} \textbf{Param.} & \textbf{Val.}& \cellcolor[HTML]{EFEFEF} \textbf{Param.} & \textbf{Val.}\\
\hline
 \cellcolor[HTML]{EFEFEF}  $N$  & $100$ & \cellcolor[HTML]{EFEFEF} $\beta$ & $2$&
\cellcolor[HTML]{EFEFEF} $K_{\rmR}$  & $2.5$&
\cellcolor[HTML]{EFEFEF} $P_G$ & $2NM$ \\
\hline 
\cellcolor[HTML]{EFEFEF}  $\bar{\beta}_{\Los}$ & $2$ & \cellcolor[HTML]{EFEFEF} $\bar{\beta}_{\Nlos}$ & $4$ &
\cellcolor[HTML]{EFEFEF} $\bar{K}_{\rmR,\Los}$ & $2.5$ & \cellcolor[HTML]{EFEFEF} $\bar{K}_{\rmR,\Nlos}$ & $0$ \\
\hline 
\cellcolor[HTML]{EFEFEF} $\bar{P}$ & $2N$  &
\cellcolor[HTML]{EFEFEF} $K_{\rmL,\Los}$ & $2$ & \cellcolor[HTML]{EFEFEF} $K_{\rmL,\Nlos}$ & $0$  & \cellcolor[HTML]{EFEFEF} $\beta_{\rmL,\Los}$  &  $2$ \\
\hline
\cellcolor[HTML]{EFEFEF} $\beta_{\rmL,\Nlos}$ & $4$ &
\cellcolor[HTML]{EFEFEF} $M$ & $8$ &
\cellcolor[HTML]{EFEFEF} $P_{\rmL}$ & $2M$ & \cellcolor[HTML]{EFEFEF} $\sigma_n^2$ & $-80$ dBm

\end{tabular}%
}
\else
\resizebox{0.49\textwidth}{!}{%
\renewcommand{\arraystretch}{1.3}
\begin{tabular}{c|c|c|c|c|c|c|c}

\cellcolor[HTML]{EFEFEF} \textbf{Param.} & \textbf{Val.} & \cellcolor[HTML]{EFEFEF} \textbf{Param.}& \textbf{Val.} & \cellcolor[HTML]{EFEFEF} \textbf{Param.} & \textbf{Val.}& \cellcolor[HTML]{EFEFEF} \textbf{Param.} & \textbf{Val.}\\
\hline
 \cellcolor[HTML]{EFEFEF}  $N$  & $100$ & \cellcolor[HTML]{EFEFEF} $\beta$ & $2$&
\cellcolor[HTML]{EFEFEF} $K_{\rmR}$  & $2.5$&
\cellcolor[HTML]{EFEFEF} $P_G$ & $2NM$ \\
\hline 
\cellcolor[HTML]{EFEFEF}  $\bar{\beta}_{\Los}$ & $2$ & \cellcolor[HTML]{EFEFEF} $\bar{\beta}_{\Nlos}$ & $4$ &
\cellcolor[HTML]{EFEFEF} $\bar{K}_{\rmR,\Los}$ & $2.5$ & \cellcolor[HTML]{EFEFEF} $\bar{K}_{\rmR,\Nlos}$ & $0$ \\
\hline 
\cellcolor[HTML]{EFEFEF} $\bar{P}$ & $2N$  &
\cellcolor[HTML]{EFEFEF} $K_{\rmL,\Los}$ & $2$ & \cellcolor[HTML]{EFEFEF} $K_{\rmL,\Nlos}$ & $0$  & \cellcolor[HTML]{EFEFEF} $\beta_{\rmL,\Los}$  &  $2$ \\
\hline
\cellcolor[HTML]{EFEFEF} $\beta_{\rmL,\Nlos}$ & $4$ &
\cellcolor[HTML]{EFEFEF} $M$ & $8$ &
\cellcolor[HTML]{EFEFEF} $P_{\rmL}$ & $2M$ & \cellcolor[HTML]{EFEFEF} $\sigma_n^2$ & $-80$ dBm

\end{tabular}%
}
\fi
\renewcommand{\arraystretch}{1}

\end{table}

\ifsingle
\begin{figure}[t!]
    \centering
    \includegraphics[width=0.5\columnwidth]{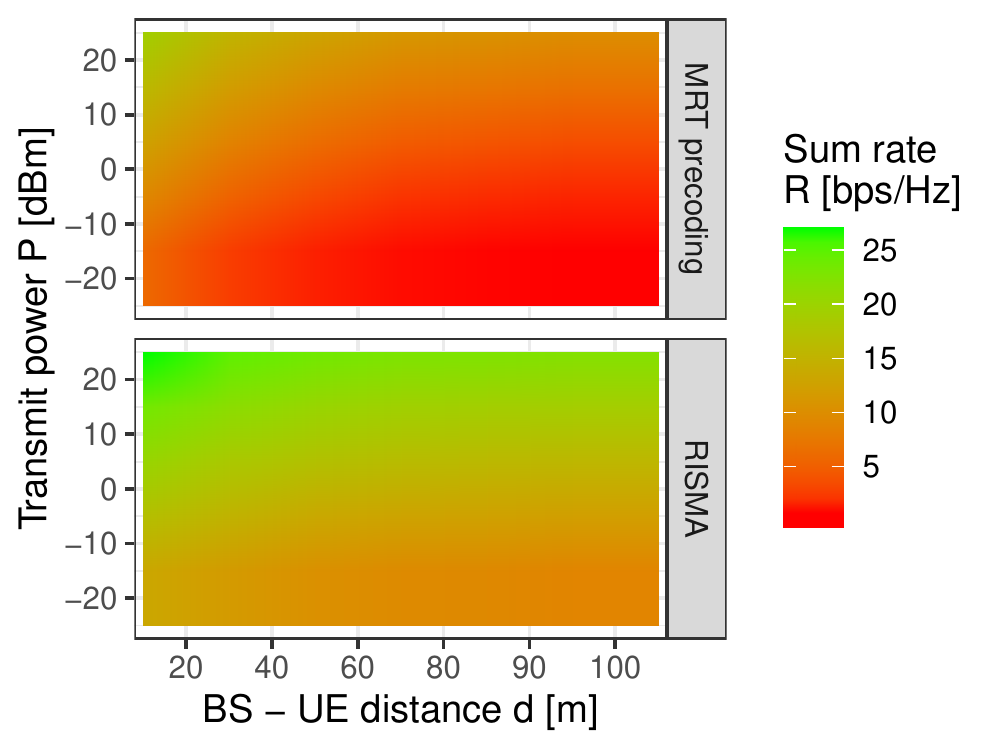}\vspace{-4mm}
    \caption{Average sum rate in the single-UE case obtained with the proposed \name{} algorithm (bottom plot) and with conventional MRT precoding (upper plot) versus the transmit power $P$ at the BS for different values of BS-UE distance $d$.}
    \label{fig:single_UE_vs_SNR_d}
\end{figure}
\else
\begin{figure}[t!]
    \centering
    \includegraphics[width=\columnwidth]{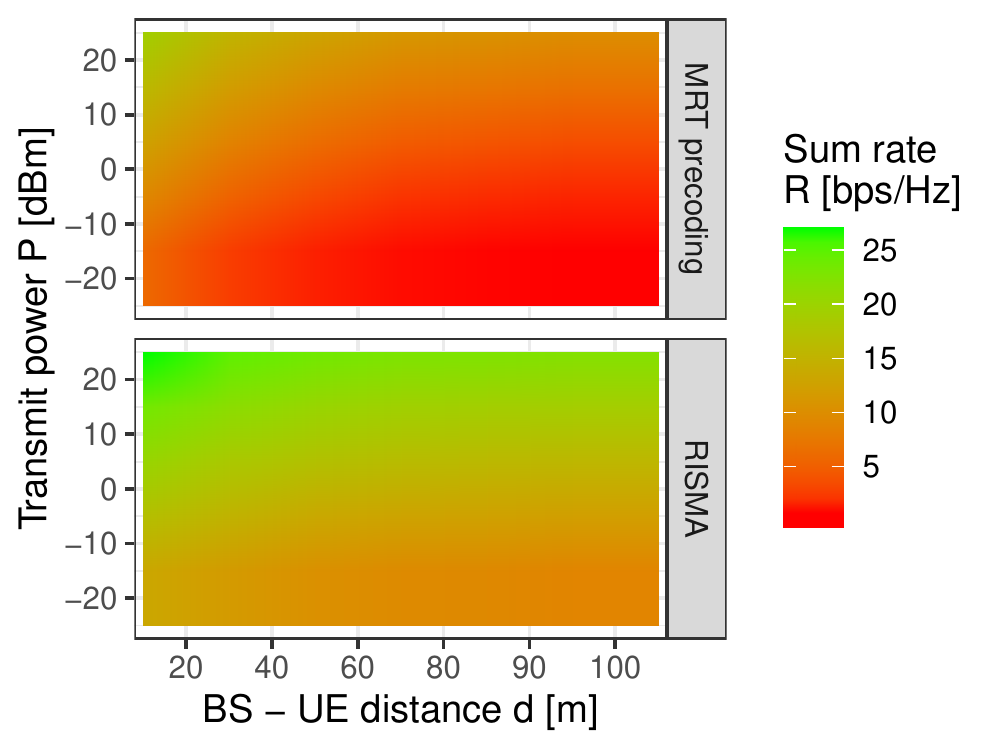}\vspace{-4mm}
    \caption{Average sum rate in the single-UE case obtained with the proposed \name{} algorithm (bottom plot) and with conventional MRT precoding (upper plot) versus the transmit power $P$ at the BS for different values of BS-UE distance $d$.}
    \label{fig:single_UE_vs_SNR_d}
\end{figure}
\fi

\subsection{Single user case}\label{subsec:Results_singleUE}
\ifsingle
\begin{figure}[!htb]
   \begin{minipage}{0.48\textwidth}
     \centering
     \includegraphics[scale=0.75]{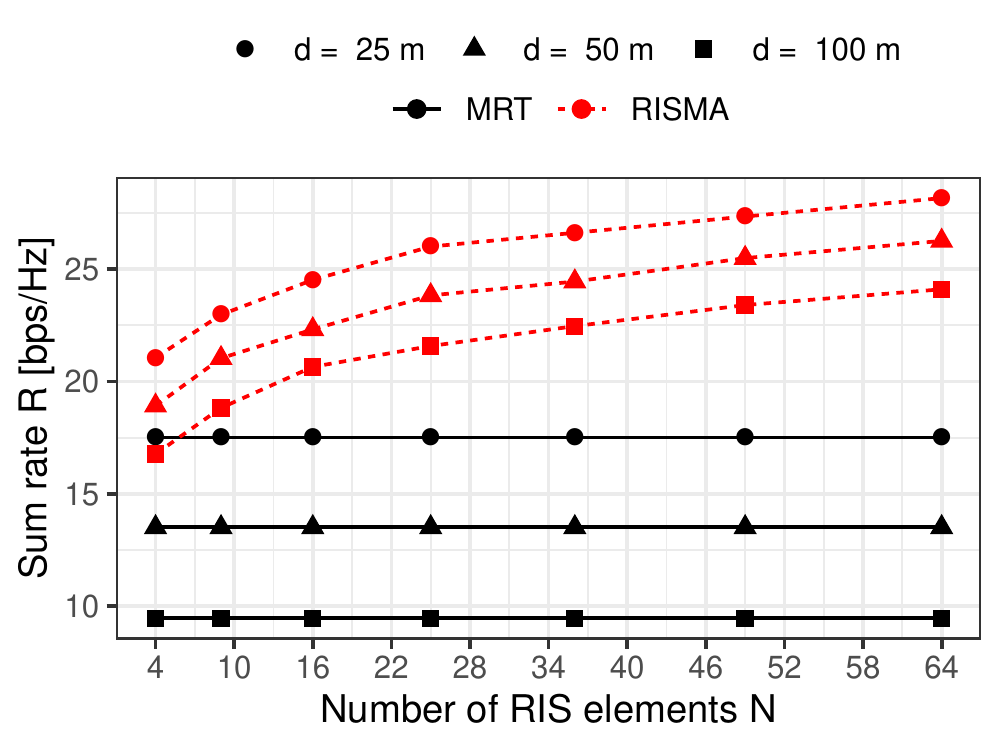}\vspace{-4mm}
    \caption{Average sum rate in the single-UE case obtained with the proposed \name{} algorithm (red lines) and with conventional MRT precoding (black lines) versus the number of RIS elements $N$ with $P=24$ dBm and for different values of the distance between the BS and the UE $d$.}
    \label{fig:single_UE_vs_N}
   \end{minipage}\hfill
   \begin{minipage}{0.48\textwidth}
     \centering
     \includegraphics[scale=0.75]{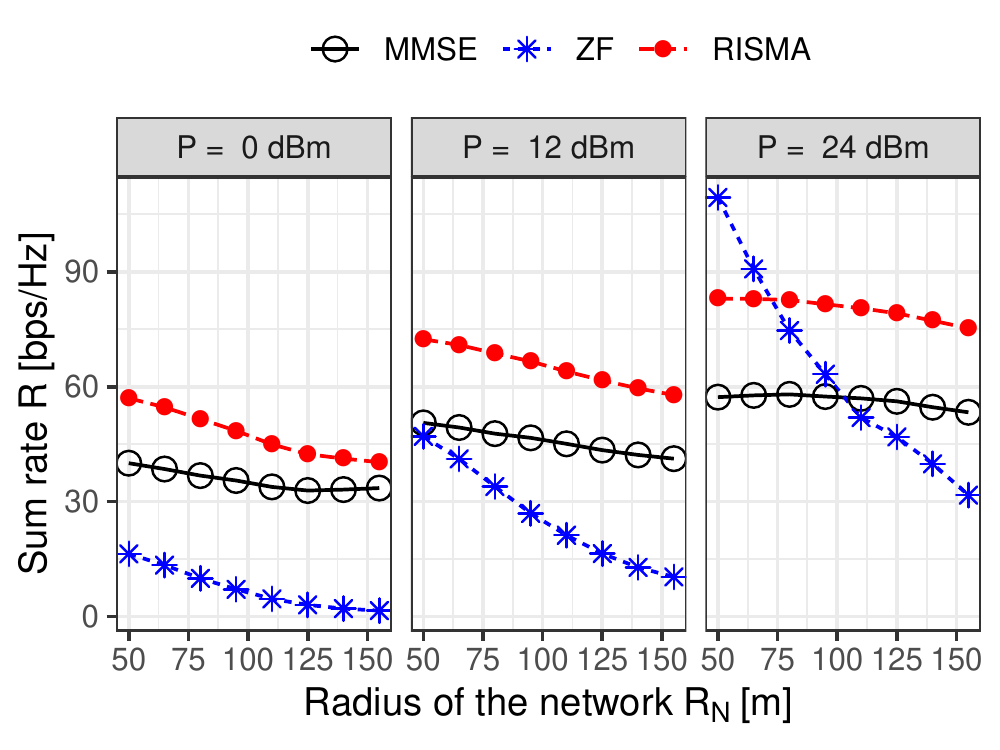}\vspace{-4mm}
    \caption{Average sum rate in the multi-UE case obtained with the proposed \name{} algorithm (red curves), with conventional MMSE precoding (black curves) and with conventional ZF precoding (red green curves) versus the radius of the network area $R_N$ and for different values of the transmit power $P$.}
    \label{fig:multi_UE_vs_Rad}
   \end{minipage}
\end{figure}
\else
\begin{figure}[t!]
    \centering
    \includegraphics[scale=0.75]{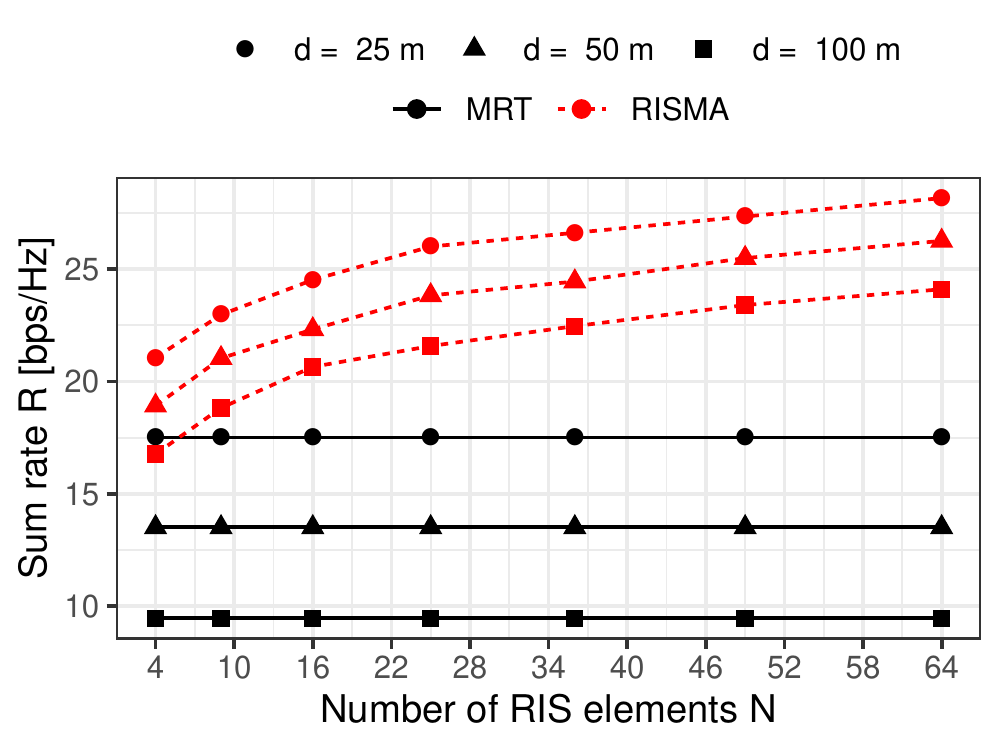}\vspace{-4mm}
    \caption{Average sum rate in the single-UE case obtained with the proposed \name{} algorithm (red lines) and with conventional MRT precoding (black lines) versus the number of RIS elements $N$ with $P=24$ dBm and for different values of the distance between the BS and the UE $d$.}
    \label{fig:single_UE_vs_N}
\end{figure}
\begin{figure}[t!]
    \centering
    \includegraphics[scale=0.75]{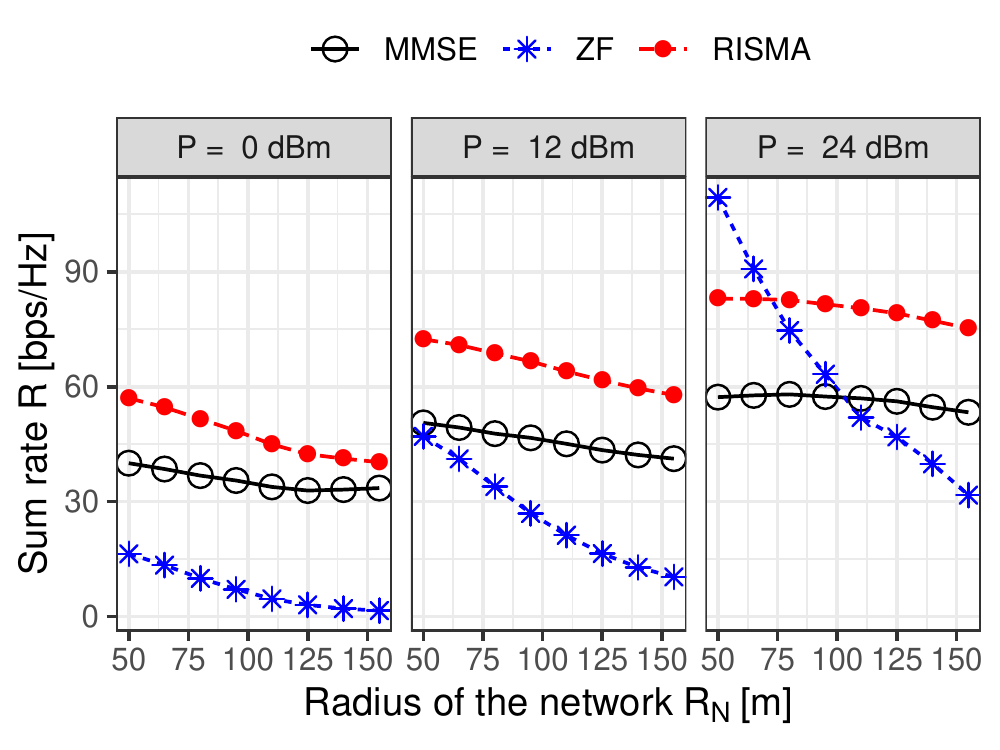}\vspace{-4mm}
    \caption{Average sum rate in the multi-UE case obtained with the proposed \name{} algorithm (red curves), with conventional MMSE precoding (black curves) and with conventional ZF precoding (red green curves) versus the radius of the network area $R_N$ and for different values of the transmit power $P$.}
    \label{fig:multi_UE_vs_Rad}
\end{figure}
\fi

In the single UE case, we set $N_x = N_y = 5$, $K_{\rmL} = 0$, $\bar{K}_{\rmR} = 2.5$, $\beta_{\rmL} = 4$, $\bar{\beta} = 2$ and $\theta = 0$. In addition, we assume that a single RIS is at distance $d_1 = 25$m from the BS with AoD $\psi_D = \pi/4$ and AoA $\psi_A = 5\pi/4$ while we vary $d$ as the distance from the BS to the UE. The distance from the RIS to the UE and the AoD $\psi$ are thus calculated based on the aforementioned parameters.

Fig.~\ref{fig:single_UE_vs_SNR_d} shows the average sum rate $R$ obtained with the proposed RIS-aided optimization in Problem \ref{eq:P3} (bottom plot) versus conventional MRT precoding (upper plot) without the aid of a RIS, defined as in Eq.~\eqref{eq:mrt} with $\Phib = \0$  versus the transmit power $P$ at the BS and for different values of distance $d$ from the BS to the UE. While both schemes exhibit an increasing sum rate with the transmit power $P$, our proposed scheme consistently outperforms a conventional network with no RIS. Moreover, the gain further increases with $d$ demonstrating how the effectiveness of a RIS in increasing the coverage area of wireless networks. For example \name{} can support up to $d=90$ m with a target sum rate of $20$ bps/Hz, thus increasing the coverage area of about $350\%$ with respect to conventional MRT precoding with no RIS. Indeed, as the UE moves far away from the BS, conventional MRT suffers from the diminishing power in the channel $\h_{\rmL}$. On the contrary the RIS-aided scheme is able to counteract this effect by steering the transmitted signal upon the RIS and towards the UE.

In Fig.~\ref{fig:single_UE_vs_N} we show the average sum rate $R$ obtained with the proposed RIS-aided optimization (red lines) versus conventional MRT (black lines) with $P = 24$ dBm versus the number of RIS elements $N$ and for different values of the BS-UE distance $d$. While the benefits introduced by adding antenna elements on the RIS are clear, interestingly the conventional MRT scheme without the aid of the RIS attains similar performance to the proposed RIS-aided scheme only for small values of $N$ and $d$. This further supports our claim that RISs can be effectively used to increase the coverage area of wireless networks at low expenses, i.e., limited number of antenna elements.

\ifsingle

    \begin{figure}[t!]
     \centering
     \includegraphics[width=\columnwidth]{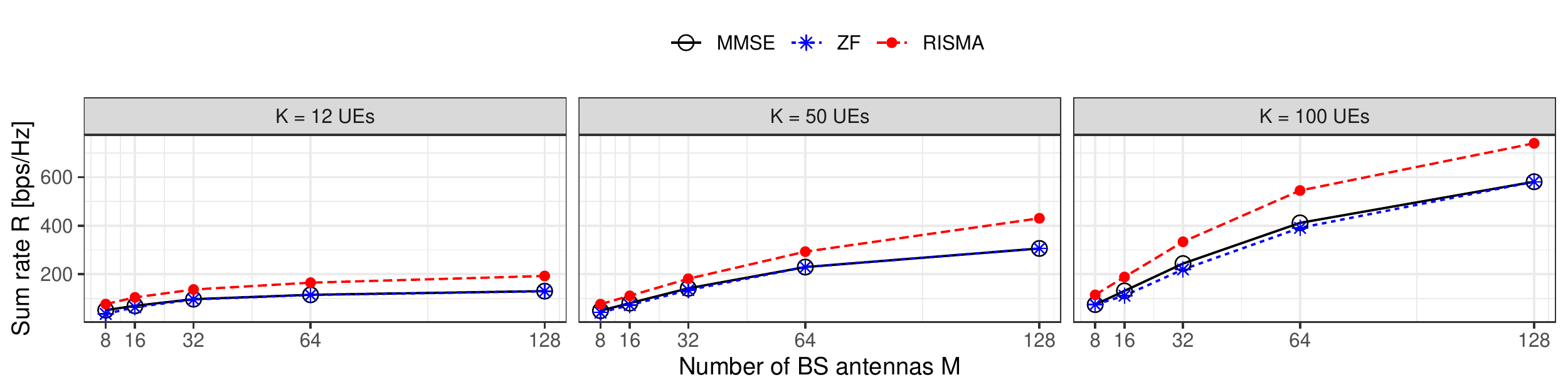}\vspace{-4mm}
    \caption{Average sum rate in the multi-UE case obtained with the proposed \name{} algorithm (red line), with conventional MMSE precoding (black line) and with conventional ZF precoding (blue line) versus the number of BS antennas $M$, for fixed network area radius $R_N=150$ m, transmit power $P = 24$ dBm and for different number of UEs $K$.}
    \label{fig:multi_UE_vs_M}
    \end{figure}
    \begin{figure}
    \begin{minipage}{0.48\textwidth}
     \centering
     \includegraphics[scale=0.75]{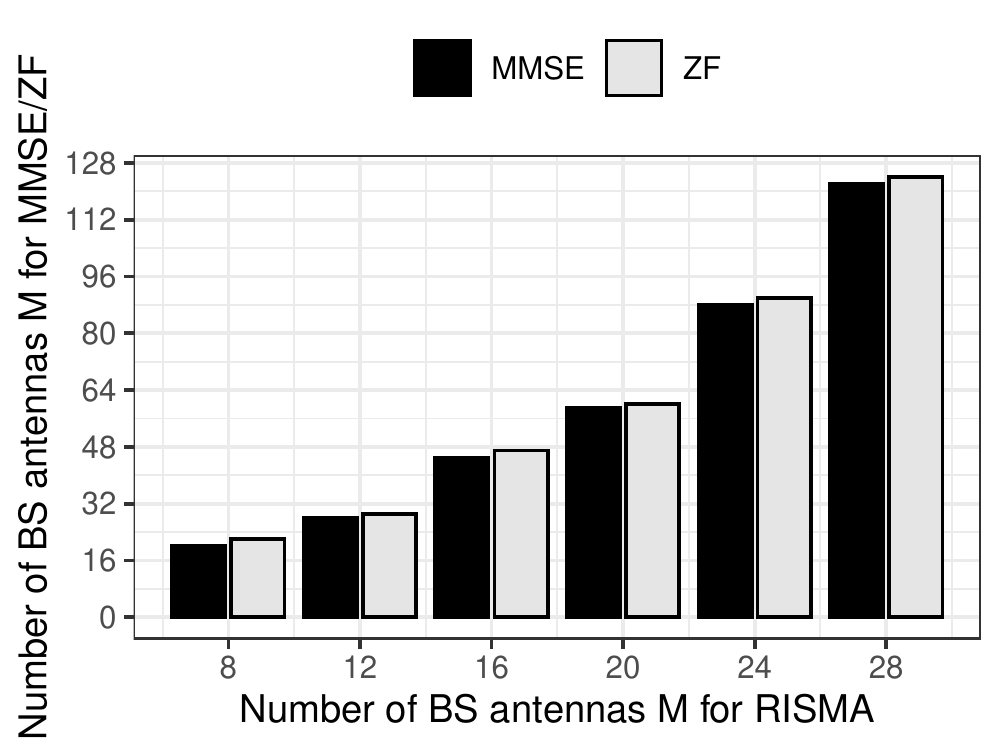}\vspace{-4mm}
    \caption{Number of BS antennas $M$ required by the MMSE (black) and ZF (blue) precoding schemes versus the number of BS antennas required by the proposed \name{} algorithm for the same targe average sum rate, fixed network area radius $R_N=100$ m and transmit power $P = 24$ dBm.}
    \label{fig:multi_UE_vsM_bars}
   \end{minipage}\hfill
   \begin{minipage}{0.48\textwidth}
    \centering
    \includegraphics[scale=0.75]{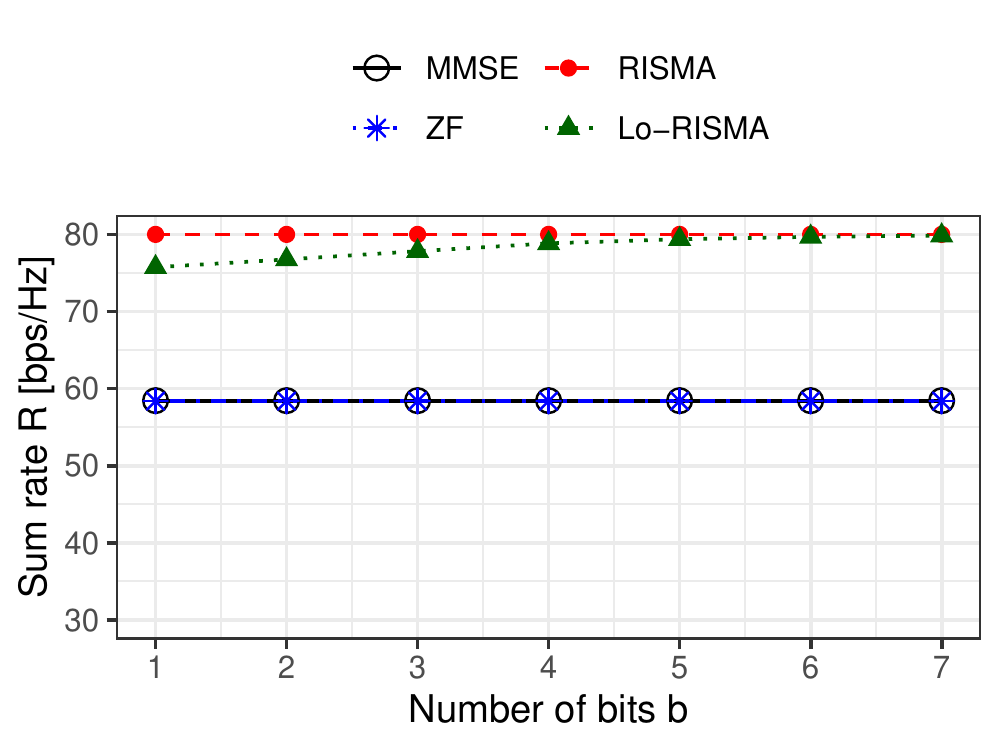}\vspace{-4mm}
    \caption{Average sum rate in the multi-UE case obtained with the new Lo-\name{} algorithm (green line), with the novel \name{} algorithm (red line), with conventional MMSE precoding (black line) and with conventional ZF precoding (blue line) versus the number of quantization bits $b$, for fixed network area radius $R_N=100$ m and transmit power $P = 24$ dBm.}
    \label{fig:multi_UE_quantized}
   \end{minipage}
   
\end{figure}
\else
\begin{figure}[t!]
    \centering
    \includegraphics[scale=0.75]{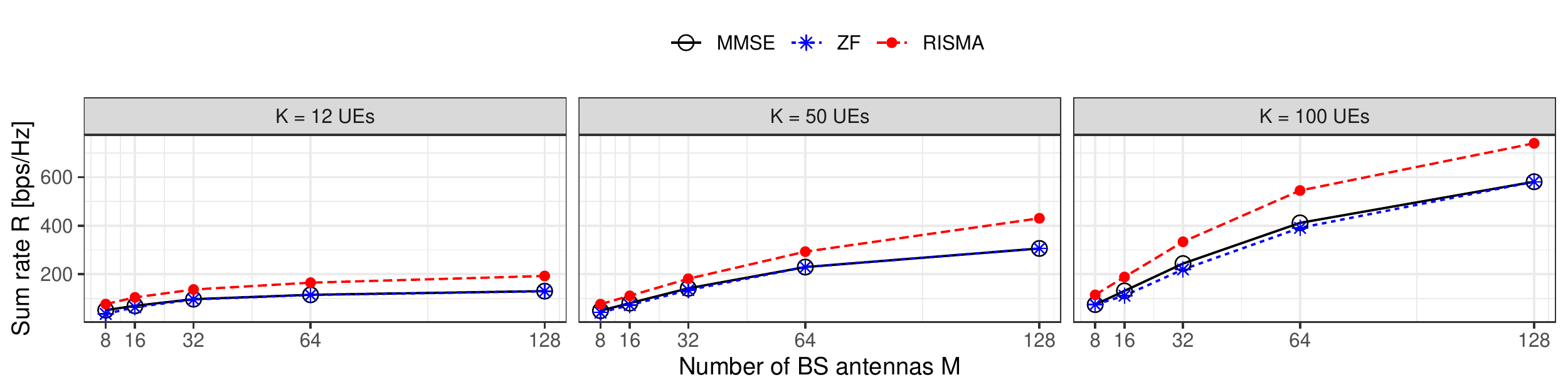}\vspace{-4mm}
    \caption{Average sum rate in the multi-UE case obtained with the proposed \name{} algorithm (red line), with conventional MMSE precoding (black line) and with conventional ZF precoding (blue line) versus the number of BS antennas $M$, for fixed network area radius $R_N=150$ m, transmit power $P = 24$ dBm and for different values of the number of UEs $K$.}
    \label{fig:multi_UE_vs_M}
\end{figure}
\begin{figure}[t!]
    \centering
    \includegraphics[scale=0.75]{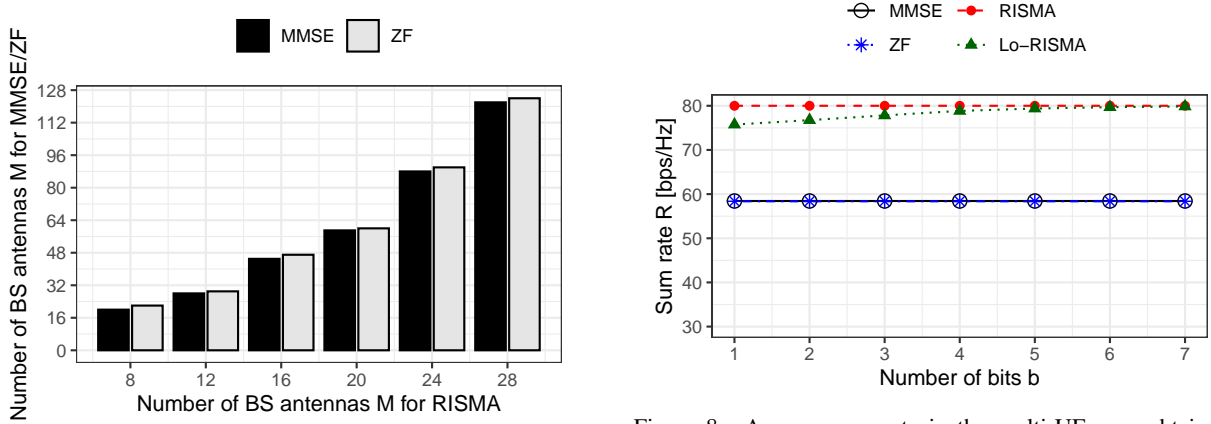}\vspace{-4mm}
    \caption{Number of BS antennas $M$ required by the MMSE (black) and ZF (blue) precoding schemes versus the number of BS antennas required by the proposed \name{} algorithm for the same targe average sum rate, fixed network area radius $R_N=100$ m and transmit power $P = 24$ dBm.}
    \label{fig:multi_UE_vsM_bars}
\end{figure}
\begin{figure}
     \centering
    \includegraphics[scale=0.75]{./img/multiUE_quantized_FINA_v2L.pdf}\vspace{-4mm}
    \caption{Average sum rate in the multi-UE case obtained with the new Lo-\name{} algorithm (green line), with the novel \name{} algorithm (red line), with conventional MMSE precoding (black line) and with conventional ZF precoding (blue line) versus the number of quantization bits $b$, for fixed network area radius $R_N=100$ m and transmit power $P = 24$ dBm.}
    \label{fig:multi_UE_quantized}
\end{figure}
\fi

\subsection{Multiuser case}\label{subsec:Results_multiUE}

In the multi-UE case we fix $K=12$ unless otherwise stated while other simulation parameters are set as per Table~\ref{tab:channel_param}. The values of $\{d_k\}_{k=1}^K$, $\{\theta_k\}_{k=1}^K$, $\{d_{2,k}\}_{k=1}^K$ and $\{\psi_k\}_{k=1}^K$ are calculated based on the UE positions of each random realization. We compare our proposed algorithms against two benchmark schemes, namely MMSE precoding defined as per~\cite{Pee05}:
\begin{align}
 \W_{\mathrm{MMSE}} = \sqrt{P} \frac{\bigg(\H_{\rmL}\H_{\rmL}^{\herm} + \frac{M\sigma_n^2 }{P} \I_M\bigg)^{-1}\H_{\rmL}}{\bigg\|\bigg(\H_{\rmL}\H_{\rmL}^{\herm} + \frac{M\sigma_n^2 }{P} \I_M\bigg)^{-1}\H_{\rmL}\bigg\|_{\mathrm{F}}},   
\end{align}
with $\H_{\rmL} = [\h_{\rmL,1},\ldots,\h_{\rmL,K}]$ and ZF precoding defined as
\begin{align}
    \W_{\mathrm{ZF}} = \sqrt{P} \frac{\H_{\rmL}\big(\H_{\rmL}^{\herm}\H_{\rmL}\big)^{-1}}{\|\H_{\rmL}\big(\H_{\rmL}^{\herm}\H_{\rmL}\big)^{-1}\|_{\mathrm{F}}}.
\end{align}

Fig.~\ref{fig:multi_UE_vs_Rad} shows the avegage sum rate $R$ in the multi-UE case obtained with the proposed \name{} algorithm (solid red line), with conventional MMSE precoding (dashed black line) and ZF precoding (dashed green line) versus the radius of the network area $R_N$ and for different values of the transmit power $P$. While the ZF scheme obtains the best results for small cells and high transmit power, the proposed \name{} algorithm achieves higher throughput for cells larger than say $75$m or a low-to-moderate transmit power. This is because as the radius of the network area increases, it becomes infeasible to design orthogonal beamformers with ZF precoding if multiple UEs are close to each other and far from the BS. Indeed, according to the proposed model, the channel response $\h_{\rmL,k}$ depends essentially on the AoD $\theta_k$ and distance $d_k$, plus the NLoS components, which carry significantly less power. Hence, UEs who are close to each other exhibit statistically similar channels, thus limiting the performance of ZF precoding due to interference among neighboring UEs. In contrast, the use of the RIS greatly alleviates interference even for closely spaced UEs in large networks thanks to steering operated by the RIS. Additionally, \name{} outperforms MMSE precoding over all the considered range of network radii and transmit powers due to the coherent sum of both the contribution of the direct path from the BS to the UE and the contribution steered by the RIS towards the intended UE. 

In Fig.~\ref{fig:multi_UE_vs_M}, for a fixed network area radius $R_N = 150$ m and transmit power $P = 24$ dBm, we vary the number of BS antennas $M$ and compare the proposed \name{} algorithm against both MMSE and ZF precoding in terms of sum rate. We evaluate three different scenarios by increasing the number of simultaneous UEs up to $K=100$, which reasonably unveils a classical IoT environment. Note that the gain brought by adding BS antennas is larger in \name{} algorithm than the two considered benchmarks since they are limited by interference due to neighboring UEs. In addition, the proposed method benefits in terms of sum rate from an increase in the number of UEs while maintaining a simple and scalable optimization routine, demonstrating its relevance in IoT scenarios. Moreover, \name{} is considerably more energy-efficient. This is made evident in Fig.~\ref{fig:multi_UE_vsM_bars} that shows the number of equivalent antennas needed by either our MMSE and ZF benchmarks to achieve the same sum gain performance than our approach with $M=\{8, 12, 16, 20, 24, 28\}$ antennas. For instance, given a target sum rate equal to $100$ bps/HZ, \name{} requires a number of BS antennas that is $\sim 67\%$ lower.

Lastly, Fig.~\ref{fig:multi_UE_quantized} shows the average sum rate $R$ obtained with the proposed Lo-\name{}, i.e., when the RIS is a low resolution metasurface, the proposed \name{} algorithm, i.e., with an ideal RIS, conventional MMSE and ZF precoding versus the number of quantization bits $b$ for fixed network area radius $R_N = 100$ m and transmit power $P = 24$ dBm. As the number of quantization bits increases, the Lo-\name{} algorithm approaches the performance of the ideal \name{} algorithm. Moreover, even for a single bit quantization our proposed methods achieves better performance than the considered benchmark schemes thus demonstrating the feasibility of RIS-aided networks. 

Remarkably, both the proposed \name{} and Lo-\name{} algorithms converge within few iterations, specifically, between $3$ to $10$ iterations. Note that the observed lower limit in the number of iterations is due to the random initialization of the optimization variables $\W$ and $\v$ in both the proposed algorithms.

\section{Conclusions}\label{sec:concl}

In this work RIS-aided beamforming solutions have been proposed, \name{} and Lo-\name{}, for addressing massive IoT access challenges in beyond 5G networks. In particular, we have analyzed RISs benefits to cope with NLOS issues in dense urban environments where massive IoT deployments are expected in the near future.

Our contributions are: $i$) a novel mathematical framework to minimize the SMSE of RIS-aided beamforming communication systems, $ii$) \name{}, a low-complexity scheme that finds a simple and effective solution for such systems, $iii$) Lo-\name{}, an efficient algorithm for deployments with low-resolution meta-surfaces, and $iv$) a numerical evaluation that shows substantial gains in terms of sum rate performance, i.e. $40\%$ gain over an MMSE precoder and $20\%$ to $120\%$ with respect to a ZF precoder, depending on the network radius.


\begin{appendix}

\subsection{Proof of Eq.~\eqref{eq:v_opt_AO}}\label{ap:SMSE_v}
Problem~\ref{eq:P_SMSE_v} is convex and thus the optimal solution solves the KKT conditions. Let the Lagrangian and its gradient as
\ifsingle
\begin{align}
    \Lc(\v,\mub,\nu) & = \sum_k \|\v^{\herm} \bar{\H}_k \W\|^2 \!-\! 2 \sum_k \Re\{\v^{\herm}\bar{\H}_k \w_k\} \!+\! \sum_{i=1}^N \mu_i (|v_i|^2-1) + \nu \, (v_{N+1}\! -\! 1)\\
    \nabla \Lc(\v,\mub,\nu) & = \sum_k \bar{\H}_k \W \W^{\herm}\bar{\H}_k^{\herm}  \v - \sum_k \bar{\H}_k \w_k + \sum_{i=1}^N \mu_i \e_i\e_i^{\tran} \v + \nu \, \e_{N+1}
\end{align}
\else
\begin{align}
    \Lc(\v,\mub,\nu) & = \sum_k \|\v^{\herm} \bar{\H}_k \W\|^2 - 2 \sum_k \Re\{\v^{\herm}\bar{\H}_k \w_k\} \nonumber \\
    & \phantom{= }+ \sum_{i=1}^N \mu_i (|v_i|^2-1) + \nu \, (v_{N+1}-1)\\
    \nabla \Lc(\v,\mub,\nu) & = \sum_k \bar{\H}_k \W \W^{\herm}\bar{\H}_k^{\herm}  \v - \sum_k \bar{\H}_k \w_k \nonumber \\
    & \phantom{= }+ \sum_{i=1}^N \mu_i \e_i\e_i^{\tran} \v + \nu \, \e_{N+1}
\end{align}
\fi
respectively. Note that to find the derivative of the real part of $\v^{\herm}\hat{\H}_k\e_k$ we have used again the property in Eq.~\eqref{eq:f_partial_z}. The KKT conditions of Problem~\ref{eq:P_SMSE_v} can be written as
\begin{align}
& \bigg(\sum_k \bar{\H}_k \W \W^{\herm}\bar{\H}_k^{\herm} + \mathrm{diag}(\mub)\bigg) \v = \bigg(\sum_k \bar{\H}_k \w_k - \nu \,   
\e_{N+1}\bigg) \\
& |v_i|^2 \leq 1 \quad i=1,\ldots,N;\quad v_{N+1} = 1; \quad
 \mub \geq \0; \quad \mu_i (|v_i|^2 -1) = 0 \quad i=1,\ldots,N
\end{align}
whose solution is given by
\ifsingle
\begin{align}
    \v & = \bigg(\sum_k \bar{\H}_k \W \W^{\herm}\bar{\H}_k^{\herm} + \mathrm{diag}(\mub)\bigg)^{-1} \bigg(\sum_k \bar{\H}_k \w_k - \nu \, \e_{N+1}\bigg),
\end{align}
\else
\begin{align}
    \v & = \bigg(\sum_k \bar{\H}_k \W \W^{\herm}\bar{\H}_k^{\herm} + \mathrm{diag}(\mub)\bigg)^{-1} \nonumber \\
    & \times \bigg(\sum_k \bar{\H}_k \w_k - \nu \, \e_{N+1}\bigg),
\end{align}
\fi
with $\mub \geq \0$ found in the following way
\begin{align}
    & \mu_i = 0 \text{~and~} |v_i|^2 \leq 1, \nonumber \\
    & \mu_i \geq 0 \text{~and~} |v_i|^2 = 1, \quad \forall i=1,\ldots,N.
\end{align}
Lastly, $\nu$ is determined by forcing $v_{N+1} = 1$. Note that
\begin{align}
    v_{N+1} & = \e_{N+1}^{\tran} \v \\
    & =  \e_{N+1}^{\tran} \B \, \z - \nu \, \e_{N+1}^{\tran}\B \,\e_{N+1}
\end{align}
where we have defined 
\begin{align}
    \B = \bigg(\sum_k \bar{\H}_k \W \W^{\herm}\bar{\H}_k^{\herm} + \mathrm{diag}(\mub)\bigg)^{-1},
\end{align}
and $\z = \sum_k \bar{\H}_k \w_k $. Hence we have that
\begin{align}
    \nu = \frac{\e_{N+1}^{\tran} \B \, \z - 1}{\e_{N+1}^{\tran}\B \,\e_{N+1}}.
\end{align}

\subsection{Proof of Eq.~\eqref{eq:W_opt_AO}}\label{ap:SMSE_w}
Problem~\ref{eq:P_SMSE_w} is convex and thus the optimal solution solves the KKT conditions. Let the Lagrangian and its gradient as
\ifsingle
\else
\begin{align}
    \Lc(\W,\mu) & \!=\! \|\bar{\H}^{\herm}\W\|_{\mathrm{F}}^2\! -\! 2\, \tr(\Re\{\bar{\H}^{\herm}\W\})\! +\! \mu (\|\W\|_{\mathrm{F}}^2 -P) \\
    \nabla \Lc(\W,\mu) & = \bar{\H}\bar{\H}^{\herm}\W - \bar{\H} + \mu \W
\end{align}
\fi
respectively. Note that to find the derivative of the real part of $\tr(\bar{\H}^{\herm}\W)$ we have used the following property, valid for any given scalar function $f(z)$ of complex variable $z$
\begin{align}\label{eq:f_partial_z}
    \frac{d f(z)}{d z} = \frac{1}{2}\bigg(\frac{\partial f(z)}{\partial \Re\{z\}} - j \frac{\partial f(z)}{\partial \Im\{z\}}\bigg).
\end{align}
The KKT conditions of Problem~\ref{eq:P_SMSE_w} can be written as the following
\begin{align}
    (\bar{\H}\bar{\H}^{\herm} + \mu \I_M) \W  = \bar{\H}; 
    \quad\|\W\|_{\mathrm{F}}^2 \leq P; 
    \quad\mu \geq 0; 
    \quad\mu( \|\W\|_{\mathrm{F}}^2 - P) = 0;
\end{align}
whose solution is given by $\W  = (\bar{\H}\bar{\H}^{\herm} + \mu \I_M)^{-1}\bar{\H},$
with $\mu\geq 0$ chosen such that  $\|\W\|_{\mathrm{F}}^2 = P$ (e.g., by bisection).

\end{appendix}

\addcontentsline{toc}{chapter}{References}
\bibliographystyle{IEEEtran}
\bibliography{IEEEabrv,refs}

\end{document}